\documentclass{article} % For LaTeX2e
\usepackage{iclr2026_conference,times}
\usepackage[english]{babel}
\usepackage[linesnumberedhidden,ruled,vlined, algo2e]{algorithm2e}
\usepackage[utf8]{inputenc}
\usepackage{algorithm}
\usepackage{algpseudocode}
\usepackage{amsfonts}
\usepackage{amsmath}
\usepackage{amssymb}
\usepackage{amsthm}
\usepackage{booktabs}
\usepackage{comment}
\usepackage{graphicx}
\usepackage{hyperref}
\usepackage{cleveref}
\usepackage{mathrsfs}
\usepackage{natbib}
\usepackage{pifont}
\usepackage{upgreek}
\usepackage{xcolor}
\usepackage{multirow}
\usepackage[table]{xcolor}
\usepackage[inline]{enumitem}

\newcommand{\cellred}[1]{\cellcolor{red!15}#1}
\newcommand{\cellblue}[1]{\cellcolor{blue!15}#1}
\def\rset{\mathbb{R}}
\def\rmd{\mathrm{d}}
\def\eqsp{\;}

\newcommand{\tcmrr}[1]{\textcolor{red!50}{#1}}
\newcommand{\tcmv}[1]{\textcolor{green}{#1}}
\newcommand{\tcmb}[1]{\textcolor{blue}{#1}}
\newcommand{\tcmbb}[1]{\textcolor{blue!50}{#1}}

\title{Fast, faithful and photorealistic diffusion-based image super-resolution with enhanced Flow Map models}

\author{
Maxence Noble$^{1,2}$\thanks{Work done during internship at Jasper Research.}, Gonzalo Iñaki Quintana$^{1}$, Benjamin Aubin$^{1}$  \& Clément Chadebec$^{1}$ \\
\\[-0.3em]
$^{1}$Jasper Research, Paris, France, $^{2}$CMAP, CNRS,
Ecole polytechnique, Palaiseau, France
}

\iclrfinalcopy

\begin{document}

\maketitle

\begin{figure}[h!]
    \centering
    \includegraphics[width=\linewidth]{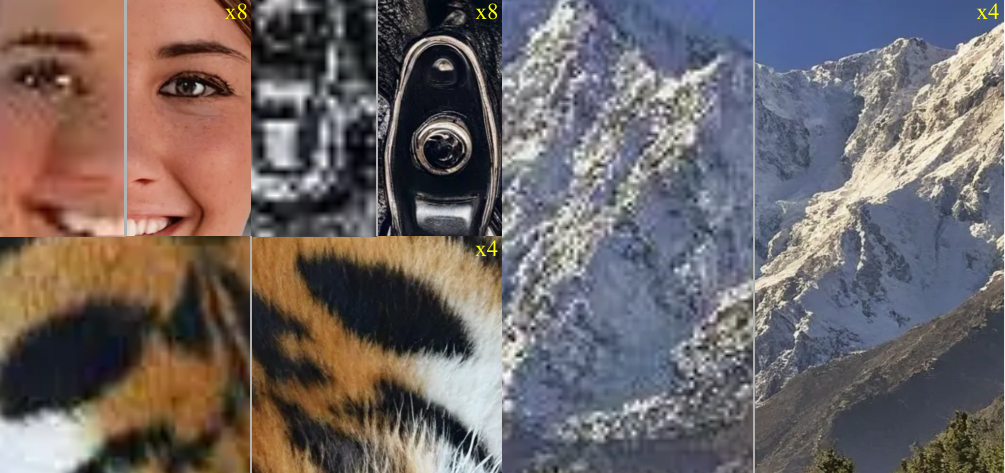}
    \vspace{-0.6cm}
    \caption{\textbf{Qualitative super-resolution results of the \emph{Shortcut} variant of FlowMapSR for $\times 4$ and $\times 8$ upscaling.} A single Flow Map model is used for both upscaling factors, with 2 inference steps.}
    \label{fig:placeholder}
\end{figure}

\begin{abstract}
    Diffusion-based image super-resolution (SR) has recently attracted significant attention by leveraging the expressive power of large pre-trained text-to-image diffusion models (DMs). A central practical challenge is resolving the trade-off between reconstruction faithfulness and photorealism. To address inference efficiency, many recent works have explored knowledge distillation strategies specifically tailored to SR, enabling one-step diffusion-based approaches. However, these teacher–student formulations are inherently constrained by information compression, which can degrade perceptual cues such as lifelike textures and depth of field, even with high overall perceptual quality. In parallel, self-distillation DMs, known as Flow Map models, have emerged as a promising alternative for image generation tasks, enabling fast inference while preserving the expressivity and training stability of standard DMs. Building on these developments, we propose FlowMapSR, a novel diffusion-based framework for image super-resolution explicitly designed for efficient inference. Beyond adapting Flow Map models to the SR task, we introduce two complementary enhancements: (i) positive–negative prompting guidance, based on a generalization of classifier free-guidance paradigm to Flow Map models, and (ii) adversarial fine-tuning using Low-Rank Adaptation (LoRA). Among the considered Flow Map formulations (\emph{Eulerian}, \emph{Lagrangian}, and \emph{Shortcut}), we find that the \emph{Shortcut} variant consistently achieves the best performance when combined with these enhancements. Extensive experiments show that FlowMapSR achieves a better balance between reconstruction faithfulness and photorealism than recent state-of-the-art methods for both $\times 4$ and $\times 8$ upscaling, while maintaining competitive inference time. Notably, a single model is used for both upscaling factors, without any scale-specific conditioning or degradation-guided mechanisms.
\end{abstract}

\newpage
\section{Introduction}

Image super-resolution (SR) is a fundamental problem in computer vision, which aims to reconstruct a plausible high-resolution (HR) image from a low-resolution (LR) input. The task is inherently ill-posed, as SR exhibits a strong \emph{one-to-many} ambiguity: the LR observation provides only partial structural information, leaving high-frequency details fundamentally underdetermined. Consequently, SR methods must navigate a delicate trade-off between reconstruction \emph{faithfulness}—recovering details consistent with the underlying HR image—and \emph{photorealism}—producing visually plausible textures for the human eye. This trade-off becomes increasingly pronounced as the upscaling factor increases, since the LR input becomes less informative and the space of plausible solutions expands, and is further complicated by unknown degradations such as blur, sensor noise, and compression artefacts.

Owing to these challenges, SR has long served as a driving force for advances in generative modeling. While early learning-based approaches were dominated by GAN-based methods \citep{wang2018esrgan,wang2021realesrgan} and Transformer architectures \citep{lu2022transformer}, recent research has increasingly focused on diffusion models (DMs) \citep{sohl-dickstein2015deep,ho2020denoising,song2021score} and related flow-matching (FM) formulations \citep{lipman2022flow}, which have demonstrated a strong capacity to synthesize fine details and realistic textures in image generation tasks. As a result, diffusion-based approaches now define the state-of-the-art for image SR. Existing diffusion-based SR approaches can be broadly divided into two categories. The first consists of inference-only methods, which leverage a pre-trained DM and apply advanced posterior sampling techniques to solve the SR inverse problem without need of ground truth LR-HR data pairs or additional training; see \cite{janati2025bridging} for a comprehensive overview. Although these methods achieve consistently strong results across a variety of SR scenarios, they typically require dozens of model evaluations per image, leading to prohibitive computational costs for large-scale applications. The second category specifically relies on LR-HR training data pairs (the LR inputs being synthetically obtained from available HR images) and focuses on training DMs explicitly for SR, either from scratch \citep{li2022srdiff,saharia2022image,yue2023resshift,yue2024efficient,luo2023image,delbracio2023inversion} or by fine-tuning large pre-trained backbones \citep{lin2024diffbir,wang2024exploiting,wu2024seesr,yang2024pixel}. Although these approaches deliver strong performance, they also rely on tens to hundreds of sampling steps at inference time.

Beyond the specific context of SR, reducing the number of inference steps has become an active research direction for DMs more broadly. Prominent strategies include the development of efficient samplers \citep{song2020denoising,lu2025dpm}, the design of straight or near-optimal transport paths \citep{liu2022flow,esser2024scaling,wang2024rectified,kim2024simple}, and various knowledge distillation techniques \citep{salimans2022progressive,sauer2024adversarial,sauer2024fast,yin2024one,chadebec2025flash}. While these approaches have proven highly effective for unconditional or text-to-image generation, they do not directly translate to SR, where generation is strongly conditioned on the input image whose informativeness depends on both the degradation process and the upscaling factor. To address this challenge, several one-step diffusion-based SR methods (i.e., requiring only a single inference step) have recently emerged, primarily relying on teacher–student distillation. These approaches notably include adversarial distillation \citep{zhang2024degradation,wu2025onestepdiffusionbasedrealworldimage,xie2024addsr,he2024one}, score distillation \citep{wu2024one}, and consistency techniques \citep{wang2024sinsr}, often combined with perceptual regularization losses. Despite their efficiency and strong empirical performance, those methods are fundamentally constrained by information loss between the teacher and student models, which may legitimately degrade photorealism, restrict texture diversity, and reduce robustness across different upscaling factors.

Recently, a new paradigm for fast diffusion inference, referred to as Flow Map models, has been introduced \citep{boffi2025buildconsistencymodellearning,boffi2025flow}, with concrete instantiations proposed in the literature \citep{geng2025meanflowsonestepgenerative,frans2025one, sabour2025alignflowscalingcontinuoustime}. These models adopt a \emph{self-distillation} paradigm: starting from a FM formulation, they learn a dynamical mapping between a source and a target distribution while enforcing self-consistency across different discretization scales through an additional training objective. This formulation enables stable training in high-dimensional spaces and supports inference with very few (potentially a single) model evaluations. Flow Map models have recently demonstrated state-of-the-art performance for image generation at resolution $512\times512$ \citep{sabour2025alignflowscalingcontinuoustime}, and ongoing work has begun exploring their scalability to higher resolution \citep{zheng2025large,hu2025cmt}. Crucially, unlike distillation approaches that compress an iterative denoising process into the student network—thereby potentially incurring information loss—Flow Map models preserve the expressivity of large generative models through self-consistency, making them particularly appealing for the SR task.

In this work, we investigate the use of Flow Map models for image SR. Our objective is to combine the expressive power of large DMs with fast inference, while maintaining both faithfulness to the underlying HR content and strong perceptual quality. Our contributions are threefold:
\begin{itemize}[wide, labelindent=0pt]
\item We introduce \textbf{FlowMapSR}, a simple and generic diffusion-based framework for SR built upon the three major Flow Map formulations (\emph{Eulerian}, \emph{Lagrangian}, \emph{Shortcut}). To adapt Flow Map models to the specific challenges of SR, we propose a base implementation that guarantees perceptual quality along with two enhancements: positive–negative prompting guidance inspired by \cite{zhang2024degradation}, and lightweight adversarial fine-tuning using LoRA modules \citep{hu2022lora}.
\item We empirically demonstrate that the FlowMapSR instance based on the \emph{Shortcut} formulation \citep{frans2025one} (equivalently the \emph{Progressive} approach from \cite{boffi2025buildconsistencymodellearning}) provides the best performance and scalability in high-resolution SR scenarios. With inference time comparable to existing one-step state-of-the-art SR methods, this version of FlowMapSR produces reconstructions that are in general more faithful to the HR reference while exhibiting improved photorealistic characteristics, including realistic textures, depth of field, and high-frequency details. We validate these results on both synthetic and real-world benchmarks, and demonstrate consistent performance across challenging upscaling scenarios ($\times 4$ and $\times 8$) using a single model, without additional degradation-specific guidance or conditioning.
\item Finally, we present a comprehensive ablation study of the proposed framework. Our analysis highlights a clear trade-off between reconstruction faithfulness and perceptual realism, providing insights into how FlowMapSR can be tuned to different application requirements.
\end{itemize}

\section{Related works}

Diffusion models have emerged as the dominant paradigm for image SR, thanks to their strong generative capabilities and their ability to overcome limitations of earlier GAN-based approaches \cite{wang2018esrgan, wang2021realesrgan}, that were prone to training instability and lack of photorealism. Early diffusion-based SR methods such as SR3 \citep{saharia2022image} and SRDiff \citep{li2022srdiff} adapted the conditional DDPM framework \citep{ho2020denoising} by explicitly conditioning the generation process on the LR input in pixel space. IDM \citep{gao2023implicit} further extends this line of work by incorporating an implicit neural representation, enabling SR across arbitrary continuous upscaling factors. Thereafter, several methods extended diffusion-based SR to latent space by leveraging large pre-trained text-to-image backbones, such as Stable Diffusion \citep{rombach2022high}. Notable examples include DiffBIR \citep{lin2024diffbir}, StableSR \citep{wang2024exploiting}, and SeeSR \citep{wu2024seesr}, which exploit rich semantic and perceptual representations learned at scale to improve visual quality. Despite their scalability and improved semantic modeling, these approaches still rely on the standard iterative diffusion inference process, and consequently require a large number of inference steps—typically more than 50. To attenuate this issue, some methods such as ResShift \citep{yue2023resshift,yue2024efficient}, IR-SDE \citep{luo2023image}, and InDI \citep{delbracio2023inversion} accelerated inference by designing shorter Markov chains or optimizing transition schedules, cutting the number of inference steps from hundreds to a dozen with minimal loss in reconstruction quality.

A complementary and increasingly active research direction investigates one-step (or few-step) diffusion-based formulations via distillation techniques. In this setting, a large pre-trained text-to-image DM acts as a teacher and guides the training of a lightweight student network, enabling SR with a drastically reduced number of model evaluations. Existing methods primarily differ in the choice of distillation objective and the regularization mechanisms used to preserve fidelity and perceptual quality. We may distinguish:
\begin{enumerate*}[label=\textbf{(\alph*)}]
\item \textbf{consistency-based distillation}, in which the student is trained to match the teacher’s outputs through direct regression or bidirectional consistency constraints, see for instance SinSR \citep{wang2024sinsr};
\item \textbf{score distillation}, which aims to minimize a predefined divergence between the teacher and student output distributions. OSEDiff \citep{wu2024one} follows this approach and incorporates the LPIPS perceptual loss \citep{zhang2018unreasonable} to encourage the recovery of high-frequency details;
\item \textbf{adversarial distillation}, which explicitly aligns the student’s output distribution with that of the teacher using an auxiliary discriminator. For example, ADDSR \citep{xie2024addsr} introduces a timestep-adaptive adversarial loss that dynamically balances teacher–student alignment across diffusion steps, leading to improved perceptual realism. Similarly, TAD-SR \citep{he2024one} combines a time-aware adversarial loss with high-frequency-enhanced score distillation to better recover fine textures. Along the same line, S3Diff \citep{zhang2024degradation} enhances adversarial distillation by integrating LPIPS regularization, positive–negative prompting guidance, and LoRA-based degradation-aware conditioning, significantly improving robustness to real-world degradations. Finally, VPD-SR \citep{wu2025onestepdiffusionbasedrealworldimage} augments adversarial distillation with semantic-aware supervision and high-frequency perceptual losses, leveraging pre-trained vision–language representations to further improve semantic alignment and texture fidelity.
\end{enumerate*}

\section{Preliminaries on Flow Map models}\label{sec:flow_map_theory}

Consider two probability distributions $\pi_0$ (target) and $\pi_1$ (source) supported on $\rset^d$, for some $d\geq 1$. We assume access to a training dataset $\mathcal{D}=\{(x_0^i, x_1^i)\}_{i=1}^N$, consisting of $N$ paired samples drawn from a joint coupling of $\pi_0$ and $\pi_1$, denoted by $\pi_{0,1}$. In this section, we study the problem of generating new samples from $\pi_0$ based on \textit{unseen} samples from $\pi_1$. 

A natural approach to this conditional generative modeling problem is to learn a transport map $\mathrm{T}:\rset^d \to \rset^d$ that pushes $\pi_1$ to $\pi_0$ (such that $\mathrm{T}(x_1)\sim\pi_0$ whenever $x_1\sim \pi_1$) by leveraging the available samples from $\mathcal{D}$. In practice, this is commonly addressed within the Flow Matching (FM) \citep{lipman2022flow} setting, which we briefly recall in \Cref{subsec:flowmatching}. Basically, FM models learn the drift of an Ordinary Differential Equation (ODE) whose solution transports $\pi_1$ to $\pi_0$, using a neural network. The resulting dynamics (neural ODE) defines a valid candidate for $\mathrm{T}$, by composing a sequence of local transport maps obtained by numerically integrating the neural ODE over small time intervals.

While this approach is conceptually simple and has proven effective across a wide range of data modalities, it comes with a significant computational cost at inference time. In particular, FM inference generally requires dozens of neural network evaluations to control time discretization error, which can make FM-based methods impractical in large-scale settings. Flow Map models \citep{boffi2025buildconsistencymodellearning, boffi2025flow} extend the FM framework by explicitly addressing this limitation. Instead of relying on ODE numerical integration, they introduce an additional self-distillation objective which aims to learn \emph{exact} integrations of the FM-based neural ODE on larger time intervals, thus enabling efficient inference with substantially fewer evaluations. In \Cref{subsec:flowmap}, we review the main components of this recent approach. We then introduce in \Cref{subsection:cfg_flowmap} additional elements on conditioned Flow Map models that will serve as the foundation of our FlowMapSR framework.

\subsection{Background on Flow Matching} \label{subsec:flowmatching}

\paragraph{Interpolant \& ODE.}FM can be seen as a particular instantiation of the Stochastic Interpolant (SI) framework \citep{albergo2023stochasticinterpolantsunifyingframework}, whose paradigm is the following : (i) defining a continuous-time path of probability distributions $(p_t)_{t\in [0,1]}$ that interpolates between the boundary conditions $p_0=\pi_0$ and $p_1=\pi_1$, and (ii) constructing a marginal-preserving ODE by identifying a velocity field $v:[0,1]\times \rset^d \to \rset^d$ such that the stochastic process $(X_t)_{t\in [0,1]}$ governed by
\begin{align} \label{eq:ODE}
    \rmd X_t = v_t(X_t) \rmd t\eqsp, \quad X_1\sim\pi_1 \eqsp,
\end{align}
verifies $X_t\sim p_t$ for any $t\in [0,1]$. Then, one may generate samples from $\pi_0$ by simulating ODE \eqref{eq:ODE} backward in time, starting from samples drawn from $\pi_1$. In practice, the bridging distributions $(p_t)_{t\in [0,1]}$ are defined as the marginals of a \textit{linear} interpolant $(\mathrm{I}_t)_{t\in [0,1]}$ of the form $\mathrm{I}_t = \alpha_t X_0 + \beta_t X_1$, where $(X_0,X_1)\sim \pi_{0,1}$ and $\alpha, \beta : [0,1] \to [0,1]$ are continuously differentiable functions satisfying boundary conditions $\alpha_0=\beta_1=1$ and $\alpha_1=\beta_0=0$. The standard FM setting is recovered by taking $\alpha_t=1-t$ and $\beta_t=t$. \emph{Throughout the main paper, we restrict our attention to this standard formulation}, and defer the general case to \Cref{subsec:flowmap_general}. In this setting, a valid—though generally intractable—expression for the velocity field is $v_t(x)= \mathbb{E}[v^{\text{target}}_{\text{FM}}|\mathrm{I}_t=x]$, where $v^{\text{target}}_{\text{FM}}= X_1-X_0$ corresponds to the conditional target velocity.

\paragraph{Training \& Inference.} If $v_t$ were known, a valid transport map pushing $\pi_1$ to $\pi_0$ would be given for any $x\sim \pi_1$ by 
\begin{align}\label{eq:map_theta}
    \mathrm{T}_{\text{FM}}(x) = x - \int_{0}^1 v_r(X^{1,x}_r) \rmd r \eqsp ,
\end{align}
where $(X^{t,x}_r)_{r\in [0,t]}$ denotes for any $t\in [0,1]$ the solution to ODE \eqref{eq:ODE} on $[0,t]$, with terminal condition $X_t=x$. In practice, $v_t$ has no closed form but it can still be efficiently learned by learning a parametric model $v_t^\theta$ (e.g., a neural network) via the mean-square regression problem
\begin{align}\label{eq:fm_loss}
    \arg\min_{\theta} \mathbb{E}[\mathcal{L}_{\text{FM}}(v_t^\theta, \mathrm{I}_t)] \eqsp , \eqsp 
    \mathcal{L}_{\text{FM}}(v_t^\theta, \mathrm{I}_t) = \|v^\theta_t(\mathrm{I}_t)-v^{\text{target}}_{\text{FM}}\|^2 \eqsp,
\end{align}
where the expectation is taken over $(X_0,X_1)\sim \pi_{0,1}$ (approximated by samples from $\mathcal{D}$) and $t\sim q(t)$, $q$ being a predefined time distribution on $[0,1]$. At inference time, replacing $v_t$ with its learned approximation $v_t^\theta$ yields a neural ODE whose associated transport map remains intractable and must itself be approximated numerically. The standard approach relies on discretizing the integral in \eqref{eq:map_theta} on small time intervals (backward in time) with numerical integration methods, starting from $t=1$. Whether using the first-order Euler scheme or high-order ODE solvers such as Heun or Runge–Kutta methods, this inference procedure induces a large number of neural network evaluations, which can imply a prohibitive cost for large scale applications. Finally, the FM interpolation framework allows one to derive an \textit{approximate} $x_0$-prediction given a state $(x_t,t)$ of the ODE trajectory, if we assume that the target velocity field is straight and constant :
\begin{align}\label{eq:x0_pred_fm}
    \hat{x}_0(x_t)=x_t - t\, v_t^\theta(x_t) \eqsp .
\end{align}
While this approximation is accurate for small time values (where it coincides with a first-order Euler expansion around $t=0$), it becomes generally loose at larger times for arbitrary couplings $\pi_{0,1}$, as it fails to capture the curvature of the neural ODE trajectory. However, in the special case where $\pi_{0,1}$ is the optimal transport plan associated with the quadratic cost between $\pi_0$ and $\pi_1$, the $x_0$-prediction \eqref{eq:x0_pred_fm} holds exactly for all $t \in [0,1]$ \citep{pooladian2023multisample, tong2023improving}. This observation has motivated a substantial body of work on few-step (and theoretically single-step) FM models \citep{liu2022rectified, lee2024improving, kim2024simple, kornilov2024optimal}, which provide an alternative to Flow Map approaches. A detailed study of these methods is beyond the scope of this paper.

\paragraph{FM enhancing via CFG.}

When data samples are associated with an additional class information $\mathbf{c}$ (e.g., a text label), the natural extension of FM consists in learning a class-conditional velocity field $v_t(x|\mathbf{c})=\mathbb{E}[v^{\text{target}}_{\text{FM}}|\mathrm{I}_t=x,\mathbf{c}]$, which replaces $v_t$ in ODE \eqref{eq:ODE}. In image generation, however, it has been empirically observed that combining $v_t(x|\mathbf{c})$ with its unconditional counterpart $v_t(x|\emptyset)$ (that is learned in standard FM) significantly improves improves sample quality. 

This popular trick, known as \emph{Classifier-Free Guidance} (CFG) \citep{ho2022classifierfreediffusionguidance}, amounts to substitute $v_t(x|\mathbf{c})$ \underline{at inference} by the following linear combination
\begin{align}\label{eq:cfg_standard}
    v^{\text{cfg}}_t(x|\mathbf{c}) = w\,v_t(x|\mathbf{c}) + (1-w) v_t(x|\emptyset) \eqsp, 
\end{align}
where $w\in(1,\infty)$ denotes the guidance strength, recovering the canonical setting in the limit $w\to1$. In practice, increasing $w$ enhances perceptual quality but often reduces faithfulness to realistic content. The conditional velocity $v_t(x|\mathbf{c})$ is learned via a class-conditional neural network $v^\theta_t(x|\mathbf{c})$ by minimizing a conditional variant of \eqref{eq:fm_loss},
\begin{align}\label{eq:cfm_loss}
    \arg\min_{\theta} \mathbb{E}[\mathcal{L}_{\text{cFM}}(v_t^\theta, \mathrm{I}_t, \mathbf{c})] \eqsp , \eqsp \mathcal{L}_{\text{cFM}}(v_t^\theta, \mathrm{I}_t, \mathbf{c}) = \|v^\theta_t(\mathrm{I}_t|\mathbf{c})-v^{\text{target}}_{\text{FM}}\|^2 \eqsp ,
\end{align}
where the expectation is taken over $(X_0,X_1,\mathbf{c})\sim\pi_{0,1}$ and $t\sim q(t)$. To expose $v^\theta$ to class-unconditional samples, a common practice is to drop the class condition $\mathbf{c}$ in \eqref{eq:cfm_loss} with a prescribed probability during training.

\subsection{Background on Flow Map models} \label{subsec:flowmap}

\paragraph{Definition \& parameterization.} Closely related to consistency-based DMs \citep{kim2024consistency, zheng2024trajectory,heek2024multistepconsistencymodels,li2025bidirectionalconsistencymodels,kim2025generalizedconsistencytrajectorymodels}, Flow Map models \citep{boffi2025flow,boffi2025buildconsistencymodellearning} and related approaches \citep{geng2025meanflowsonestepgenerative,sabour2025alignflowscalingcontinuoustime,frans2025one} extend the FM framework to alleviate its high inference cost. Their key idea is to directly approximate the \emph{integrated} velocity field over large time intervals, thereby enabling efficient inference with only a few steps. Formally, these methods aim to estimate the so-called \emph{Flow Map} $X:[0,1]^2\times \rset^d \to \rset^d$ associated with the ODE \eqref{eq:ODE}. For any pair of timesteps $(s,t)\in [0,1]^2$ such that $s\leq t$ and any point $x\in \rset^d$, the Flow Map is defined  as
\begin{align}\label{eq:flowmap_def}
    X_{s,t}(x)= x -\int_{s}^t v_r(X_r^{t,x}) \rmd r \eqsp .
\end{align}
In particular, the transport map $x \mapsto X_{0,1}(x)=\mathrm{T}_{\text{FM}}(x)$ can be obtained with a single evaluation of the Flow Map, making this formulation especially well suited for one-step inference. Following prior work \citep{boffi2025buildconsistencymodellearning, frans2025one}, we consider in the rest of the paper an Euler-like parameterization of the Flow Map given by
\begin{align} \label{eq:param_euler}
    X_{s,t}(x) = x -(t-s) u_{s,t}(x) \eqsp, \eqsp u_{s,t}(x)=\frac{1}{t-s}\int_{s}^t v_r(X_r^{t,x}) \rmd r \eqsp ,
\end{align}
where $u_{s,t}$ denotes the \textit{average} velocity field on time interval $[s,t]$ associated to ODE \eqref{eq:ODE} conditioned on terminal point $X_t=x$, that we now aim to estimate with a neural network $u^\theta_{s,t}$. Note that the instantaneous velocity field $v_t(x)$ corresponds to a zeroth-order approximation of $u_{s,t}(x)$ in the limit regime $s\to t$, i.e., $u_{t,t}(x) = v_{t}(x)$, which establishes a straightforward bridge between FM and Flow Map frameworks. Since we already know how to learn $v_t$, see \eqref{eq:fm_loss}, the remaining challenge is to estimate $u_{s,t}$ for $s<t$.

\paragraph{Self-distillation losses.} To this end, prior works introduce three consistency-based characterizations of the Flow Map, leading to so-called \textit{Self-Distillation} (SD) objective functions. Given timesteps $s<t$ along with the FM interpolant $\mathrm{I}_t$, the SD losses take the general form
\begin{align}\label{eq:sd_loss}
    \mathcal{L}_{\text{SD}}(u_{s,t}^\theta, \mathrm{I}_t)=\|u^\theta_{s,t}(\mathrm{I}_t)-\text{sg}(u^{\text{target}}_{\text{SD}})\|^2 \eqsp ,
\end{align}
where $\text{sg}$ is the $\operatorname{stopgradient}$ operator, and $u^{\text{target}}_{\text{SD}}$ depends on the chosen characterization:
\begin{enumerate}
    \item \textit{Lagrangian} SD (LSD) loss \citep[Eq. (13)]{boffi2025buildconsistencymodellearning}
    \begin{align*}
        u_{\text{LSD}}^{\text{target}} = v^{\text{target}}_{\text{FM}} + (t-s) \partial_s u^\theta_{s,t}(\mathrm{I}_t)
    \end{align*}
    \item \textit{Eulerian} SD (ESD) loss \citep[Eq. (9)]{geng2025meanflowsonestepgenerative}, or equivalently \citep[Eq. (14)]{boffi2025buildconsistencymodellearning}
    \begin{align*}
        u_{\text{ESD}}^{\text{target}} = v^{\text{target}}_{\text{FM}}-(t-s)\left(\nabla u^\theta_{s,t}(\mathrm{I}_t) v^{\text{target}}_{\text{FM}} + \partial_t u^\theta_{s,t}(\mathrm{I}_t)\right)
    \end{align*}
    \item \textit{Shortcut} SD (SSD) loss \citep[Eq. (5)]{frans2025one}, or equivalently \citep[Eq. (15)]{boffi2025buildconsistencymodellearning}
    \begin{align*}
        u_{\text{SSD}}^{\text{target}}= \frac{1}{2}u^\theta_{r,t}(\mathrm{I}_t) + \frac{1}{2}u^\theta_{s,r}\left(\mathrm{I}_t -\frac{t-s}{2} u^{\theta}_{r,t}(\mathrm{I}_t)\right), \quad r=\frac{s+t}{2}
    \end{align*}
\end{enumerate}
For completeness, we provide detailed derivations of these objectives in \Cref{subsec:flowmap_general}. We note that SD losses are more expensive to evaluate than the standard FM loss: the LSD and ESD losses indeed require computing the derivatives of $u^\theta$ with respect to time and space (which can be efficiently implemented via $\operatorname{jvp}$ operations at the cost of an additional backward pass), while the SSD loss requires two extra evaluations of $u^\theta$.

\paragraph{Training \& Inference.} Learning the Flow Map amounts to simply combining the flow matching loss $\mathcal{L}_{\text{FM}}$ and the self-distillation loss $\mathcal{L}_{\text{SD}}$ leading to the optimization problem
\begin{align}\label{eq:flowmap_loss}
    \arg\min_{\theta} \mathbb{E}[\mathcal{L}_{\text{FM-SD}}(u_{s,t}^\theta, \mathrm{I}_t)] \eqsp , \eqsp
    \mathcal{L}_{\text{FM-SD}}(u_{s,t}^\theta, \mathrm{I}_t) = \left\{
    \begin{array}{ll}
        \mathcal{L}_{\text{FM}}(u^\theta_{t,t}, \mathrm{I}_t) & \mbox{if } s=t \eqsp , \\
        \mathcal{L}_{\text{SD}}(u^\theta_{s,t}, \mathrm{I}_t) & \mbox{else} \eqsp ,
    \end{array}
\right.
\end{align}
where the expectation is taken over $(X_0,X_1)\sim \pi_{0,1}$ and $(s,t)\sim \bar{q}(s,t)$, $\bar{q}$ being a predefined joint distribution over $\{(s,t)\in [0,1]^2:s\leq t\}$. Due to the Euler-like parameterization given in \eqref{eq:param_euler}, inference with a Flow Map model is straightforward. Given a time discretization $\{t_k\}_{k=0}^K$ of $[0,1]$ of size $(K+1)$, with $K\geq1$, $t_0=0$ and $t_K=1$, and an initial sample $x_K\sim \pi_1$, one can recursively compute the iterates $\{x_k\}_{k=0}^K$ via the update
\begin{align} \label{eq:update}
    x_{k} = x_{k+1} - \delta_k u^\theta_{t_k, t_{k+1}}(x_{k+1}) \eqsp ,
\end{align}
where $\delta_k=t_{k+1}-t_k$ denotes the $k$-th step-size. In particular, $x_0$ is approximately sampled from $\pi_0$ if $u^\theta$ accurately approximates $u$. Importantly, this formulation avoids the discretization error inherent to standard FM, allowing the number of inference steps $K$ to remain very small, i.e., $K\in \{1,2,4\}$ \citep{boffi2025buildconsistencymodellearning,sabour2025alignflowscalingcontinuoustime}. As in the FM setting, Flow Map models admit direct $x_0$-prediction at any intermediate step of the ODE trajectory; indeed, given a current state $(x_t,t)$, the corresponding Flow Map prediction is given by
\begin{align}\label{eq:x0_pred_flowmap}
    \hat{x}_0(x_t) = x_t - t\, u^\theta_{0,t}(x_t) \eqsp . 
\end{align}
Unlike the FM prediction in \eqref{eq:x0_pred_fm}, the Flow Map prediction \eqref{eq:x0_pred_flowmap} is \emph{exact} for any given timestep $t\in [0,1]$. 

\subsection{Generalization of CFG to the Flow Map setting} \label{subsection:cfg_flowmap} 

Assume now that each training sample $(x_0,x_1)\in \mathcal{D}$ is provided with an additional knowledge $\mathbf{c}$ (e.g., a class conditioning), that we wish to exploit to improve sample quality. In the FM setting, the CFG paradigm discussed in \Cref{subsec:flowmatching} has been shown to bring substantial gains in image generation tasks, which naturally motivates its use within Flow Map models, especially in the context of SR. However, its application is not straightforward, as CFG is originally formulated for instantaneous velocity fields, whereas Flow Map models operate on averaged velocity fields over time intervals. Building on the insights of \cite{geng2025meanflowsonestepgenerative} in the ESD setting, we propose to integrate CFG directly into Flow Map models at \emph{training stage}, by exploiting the connection between FM and Flow Map formulations across all three SD settings. This leads to new CFG-aware expressions for the LSD and SSD objectives. To promote model flexibility and reduce the risk of overfitting, we further propose to \emph{randomly} sample the guidance scale $w$ during training from the interval $(1, w_{\max})$, where $w_{\max} > 1$ is treated as a hyperparameter\footnote{In contrast, \cite{geng2025meanflowsonestepgenerative} adopt a deterministic training procedure with a fixed guidance scale.}. This design prevents specialization to a single guidance strength while still retaining the benefits of CFG at inference. Below, we detail the derivation of the Flow Map training objectives in the CFG setting.

\paragraph{Flow matching loss.} Given a guidance scale $w\in (1, w_{\max})$, we rewrite the loss $\mathcal{L}_{\text{cFM}}$ in \eqref{eq:cfm_loss} to directly introduce CFG at the velocity level, yielding the following objective
\begin{align}\label{eq:cfg_loss}
    \mathcal{L}_{\text{cfg-FM}}(u_{t,t}^\theta, \mathrm{I}_t, \mathbf{c}, w) = \|u^\theta_{t,t}(\mathrm{I}_t|\mathbf{c})-\operatorname{sg}(v^{\text{target}}_{\text{cfg-FM}})\|^2\eqsp , \eqsp v^{\text{target}}_{\text{cfg-FM}} = w\, v^{\text{target}}_{\text{FM}} + (1-w) u^\theta_{t,t}(\mathrm{I}_t|\emptyset) \eqsp .
\end{align}
This loss naturally extends the standard CFG formula \eqref{eq:cfg_standard} to the Flow Map setting, by leveraging the identity $v_t=u_{t,t}$. Compared to the unconditional setting, each loss evaluation now requires one extra model evaluation.

\paragraph{Self-distillation losses.} In the same spirit as in the FM setting, we derive the following CFG-based SD loss
\begin{align}\label{eq:cfgsd_loss}
    \mathcal{L}_{\text{cfg-SD}}(u_{s,t}^\theta, \mathrm{I}_t, \mathbf{c}, w)=\|u^\theta_{s,t}(\mathrm{I}_t|\mathbf{c})-\text{sg}(u^{\text{target}}_{\text{cSD}})\|^2 \eqsp ,
\end{align}
where $u^{\text{target}}_{\text{cSD}}$ still depends on the chosen characterization:
\begin{enumerate}
    \item CFG-based \textit{Lagrangian} SD (cfg-LSD) loss
    \begin{align*}
         u_{\text{cfg-LSD}}^{\text{target}} =  v^{\text{target}}_{\text{cfg-FM}}  + (t-s) \partial_s u^\theta_{s,t}(\mathrm{I}_t|\mathbf{c}) \eqsp , \eqsp v^{\text{target}}_{\text{cfg-FM}} = w\, v^{\text{target}}_{\text{FM}} + (1-w) u^\theta_{s,s}\left(\mathrm{I}_t -(t-s)u^\theta_{s,t}(\mathrm{I}_t|\emptyset )\mid\emptyset\right) \eqsp ,
    \end{align*}
    \item CFG-based \textit{Eulerian} SD (cfg-ESD) loss, already given in \citep[Eq. (17)]{geng2025meanflowsonestepgenerative}
    \begin{align*}
         u_{\text{cfg-ESD}}^{\text{target}} = v^{\text{target}}_{\text{cfg-FM}}-(t-s)\left(\nabla u^\theta_{s,t}(\mathrm{I}_t|\mathbf{c}) v^{\text{target}}_{\text{cfg-FM}} + \partial_t u^\theta_{s,t}(\mathrm{I}_t|\mathbf{c})\right) \eqsp , \eqsp v^{\text{target}}_{\text{cfg-FM}} = w\, v^{\text{target}}_{\text{FM}} + (1-w) u^\theta_{t,t}(\mathrm{I}_t|\emptyset) \eqsp ,
    \end{align*}
    \item CFG-based \textit{Shortcut} SD (cfg-SSD) loss
    \begin{align*}
        u_{\text{cfg-SSD}}^{\text{target}}= \frac{1}{2}u^\theta_{r,t}(\mathrm{I}_t|\mathbf{c}) + \frac{1}{2}u^\theta_{s,r}( \hat{\mathrm{I}}_s | \mathbf{c}), \eqsp \hat{\mathrm{I}}_s= \mathrm{I}_t -\frac{t-s}{2} u^{\theta}_{r,t}(\mathrm{I}_t|\mathbf{c}) \eqsp, \eqsp r=\frac{s+t}{2} \eqsp .
    \end{align*}
\end{enumerate}
Detailed derivations of these objectives are provided in \Cref{subsec:flowmap_general_cfg}. Relative to the unconditional setting, computing the SD loss incurs two additional model evaluations in the Lagrangian setting, one in the Eulerian setting, and none in the Shortcut setting.

\paragraph{Training \& Inference.} When combining both objectives previously established, we obtain the CFG-based objective
\begin{align*}
    \arg\min_{\theta} \mathbb{E}[\mathcal{L}_{\text{cfg-FM-SD}}(u_{s,t}^\theta, \mathrm{I}_t, \mathbf{c}, w)] \eqsp , \eqsp
    \mathcal{L}_{\text{cfg-FM-SD}}(u_{s,t}^\theta, \mathrm{I}_t, \mathbf{c}, w)= \left\{
    \begin{array}{ll}
        \mathcal{L}_{\text{cfg-FM}}(u^\theta_{t,t}, \mathrm{I}_t, \mathbf{c}, w) & \mbox{if } s=t \eqsp, \\
        \mathcal{L}_{\text{cfg-SD}}(u^\theta_{s,t}, \mathrm{I}_t, \mathbf{c}, w) & \mbox{else},
    \end{array}
\right.
\end{align*}
where the expectation is taken over $(X_0,X_1,\mathbf{c})\sim\pi_{0,1}$, $(s,t)\sim \bar{q}(s,t)$ and $w \sim \mathrm{U}(1,w_{\max})$. Analogously to standard CFG practice, we randomly drop the class condition $\mathbf{c}$ in \eqref{eq:cfg_loss} and \eqref{eq:cfgsd_loss} with a probability of $10\%$, ensuring that the drop is applied consistently across all evaluations of $u^\theta$ within a given setting. When this occurs, we additionally set $w = 1$, thereby recovering the original \emph{unconditional} FM formulation for \eqref{eq:cfg_loss} and the corresponding \emph{unconditional} SD formulation for \eqref{eq:cfgsd_loss}. We observed in early experiments that this design choice significantly improves the quality of the generated samples. One could further incorporate additional evaluations of the instantaneous velocity $u^\theta_{t,t}(\cdot|\mathbf{c})$ or $u^\theta_{s,s}(\cdot|\mathbf{c})$ when computing $v^{\text{target}}_{\text{FM}}$, at the cost of introducing an additional hyperparameter, as discussed in \cite[Appendix B.1]{geng2025meanflowsonestepgenerative} for the ESD setting. To avoid extra hyperparameter tuning, we do not explore this extension and leave it for future work.

Importantly, the proposed CFG-based Flow Map formulation avoids the inference-time cost increase typically associated with traditional CFG. In particular, it incurs the same computational cost as the unconditional setting, allowing us to benefit from CFG without additional overhead. Given an initial sample $x_K^{\mathbf{c}} \sim \pi_1$ with class condition $\mathbf{c}$, the unconditional update rule \eqref{eq:update} naturally becomes
\begin{align} \label{eq:update_cfg}
x_k^{\mathbf{c}} = x_{k+1}^{\mathbf{c}} - \delta_k u^\theta_{t_k,t_{k+1}}(x_{k+1}^{\mathbf{c}} | \mathbf{c}) \eqsp .
\end{align}
Similarly, the $x_0$-prediction defined in \eqref{eq:x0_pred_flowmap} extends to the conditional setting as
\begin{align} \label{eq:x0_pred_flowmap_c}
\hat{x}_0(x_t |\mathbf{c}) = x_t - t\, u^\theta_{0,t}(x_t|\mathbf{c}) \eqsp . 
\end{align}

\vspace{-0.3cm}

\section{Methodology} \label{sec:method}

We propose to address the SR task within the Flow Map framework. In this setting, each paired sample $(x_0, x_1) \sim \pi_{0,1}$ consists of a low-resolution (LR) image $x_1 \in \rset^{3 \times w \times h}$ and its high-resolution (HR) counterpart $x_0 \in \rset^{3 \times W \times H}$, with a fixed aspect ratio satisfying $H/h = W/w \geq 1$. We refer to this ratio as the SR ``upscaling factor'' and denote it by $s_{\text{up}}$, while its inverse $s_{\text{down}} = 1 / s_{\text{up}}$ corresponds to the ``downscaling factor''. In \Cref{subsec:base_flow_map}, we introduce a complete methodology tailored to the SR setting, which we term \emph{FlowMapSR}. To further enhance both reconstruction faithfulness and perceptual quality, we extend FlowMapSR with a positive–negative prompting strategy inspired by \cite{zhang2024degradation}, described in \Cref{subsec:pos_neg_flow_map}, as well as an additional adversarial fine-tuning stage detailed in \Cref{subsec:LoRA_flow_map}.

\vspace{-0.2cm}

\subsection{Base implementation of FlowMapSR} \label{subsec:base_flow_map}

\paragraph{Data.} 

Each training pair in our dataset $\mathcal{D}$ consists of a ground-truth HR image $x_0$ and a corresponding LR image $x_1$, synthetically generated by applying a sequence of random degradations to $x_0$, including downscaling by a factor $s_{\text{down}}$. We build on the widely used degradation pipeline of \cite{wang2021realesrgan}, which we adapt to better reflect real-world LR images similar to those shown in \Cref{fig:real_main}. In particular, we reduce the strength of noise components (e.g., Gaussian noise, gray noise, and blur) to emphasize the effects of downscaling and compression; implementation details are provided in \Cref{sec:implem_details}. To improve robustness across diverse SR scenarios and promote generalization, we further adopt the continuous SR strategy of \cite{gao2023implicit} by sampling $s_{\text{down}} \sim \mathrm{U}(0.1, 1)$ independently for each training pair. 

The resulting LR image is then resized back to the spatial resolution of $x_0$ using a randomly selected interpolation mode (nearest, area, bilinear, or bicubic), followed by additional blur and pixel-value clamping. This procedure yields an LR source $x_1$ with the same spatial dimensions as $x_0$, which is required by the Flow Map framework, while implicitly defining a continuous upscaling factor $s_{\text{up}} \sim \mathrm{U}(1, 10)$. Although this setup theoretically enables SR up to $\times 10$, we restrict our main evaluation to $\times 4$ and $\times 8$ upscaling, to specifically emphasize the benefits of FlowMapSR approach in challenging SR scenarios. Finally, to enable large-scale training and inference, we operate in latent space by encoding image pairs with a non-trainable variational encoder. As a result, both training and inference of the Flow Map models described in \Cref{sec:flow_map_theory} are conducted in latent space, with a final decoding step, denoted by $\mathbf{Dec}$, applied at inference time to recover the HR image.

\vspace{-0.2cm}
\paragraph{Time-dependent regularization with perceptual loss.} 

Following prior work \citep{wu2024one, zhang2024degradation}, we augment the standard Flow Map training objective \eqref{eq:flowmap_loss} with an additional $\operatorname{LPIPS}$ perceptual loss \citep{zhang2018unreasonable}, which aligns the decoded HR prediction with the ground-truth HR image. This regularization significantly accelerates convergence and improves the recovery of meaningful image features at inference. The resulting objective is
\begin{align*}
\mathcal{L}^{\text{reg}}_{\text{FM-SD}}(u^\theta_{s,t}, \mathrm{I}_t)= \mathcal{L}_{\text{FM-SD}}(u^\theta_{s,t}, \mathrm{I}_t) + \lambda^{\text{LPIPS}}_s \operatorname{LPIPS}(\hat{X}_0, X_0) \eqsp, \text{where} \eqsp\hat{X}_0 = \mathbf{Dec}(\mathrm{I}_t - t \, u^\theta_{s,t}(\mathrm{I}_t)) \eqsp,
\end{align*}
where $\lambda^{\text{LPIPS}} : [0,1] \to [0, \infty)$ is a time-dependent weight and $\hat{X}_0$ denotes an approximate pixel-space $x_0$-prediction from $\mathrm{I}_t$. Unlike previous approaches that use a constant $\lambda^{\text{LPIPS}}$, we deliberately design this weight to be \emph{strictly decreasing} with time, reflecting the accuracy of direct $x_0$ prediction from $\mathrm{I}_t$ under our Flow Map model:
\begin{enumerate*}[label=(\alph*)]
\item when $s = t$ (FM case), $\hat{X}_0$ corresponds to the decoded approximation in \eqref{eq:x0_pred_fm}, whose accuracy decreases as $s$ increases; accordingly, $\lambda^{\text{LPIPS}}$ is set higher for small $s$ and lower for large $s$;
\item when $s < t$ (SD case), $\hat{X}_0$ coincides with the exact prediction in \eqref{eq:x0_pred_flowmap} only for $s = 0$, but remains reliable for small $s$, motivating a similar time-dependent weighting.
\end{enumerate*}
The specific form of $\lambda^{\text{LPIPS}}$ used in our experiments is provided in \Cref{sec:implem_details}. 

\vspace{-0.2cm}
\paragraph{Dynamic loss weighting.} 

To ensure meaningful self-consistency, the FM-SD loss must be evaluated over a broad range of timestep pairs $(s,t)$. In practice, this can induce strong time-dependent variance that hampers stable optimization. To mitigate this issue, we follow \cite{boffi2025buildconsistencymodellearning}, inspired by \cite{karras2024analyzing}, and jointly learn a \emph{dynamic weighting} function $\lambda^\psi_{s,t}$, parameterized as a lightweight neural network described in \Cref{sec:implem_details}, alongside the Flow Map model $u^\theta_{s,t}$. This leads to the reweighted objective
\begin{align}\label{eq:loss_unconditional}
    \mathcal{L}(\theta,\psi)=\mathbb{E}[\mathcal{L}(u^\theta_{s,t},\mathrm{I}_t;\psi)] \eqsp , \eqsp \mathcal{L}(u^\theta_{s,t},\mathrm{I}_t;\psi)=\exp(-\lambda^\psi_{s,t})\mathcal{L}^\text{reg}_{\text{FM-SD}}(u^\theta_{s,t}, \mathrm{I}_t) + \lambda^\psi_{s,t} \eqsp .
\end{align}
At optimality, this formulation encourages loss gradients with respect to $\theta$ to have comparable magnitudes across different $(s,t)$ values \cite[Eq. (26)]{karras2024analyzing}, thereby stabilizing training.

\vspace{-0.2cm}
\paragraph{Discrete timestep sampling.}
As described in the following section, our experiments rely on models with discrete time inputs, which precludes the use of continuous-time sampling strategies for designing $\bar{q}$ \citep{boffi2025buildconsistencymodellearning}. Instead, inspired by \cite{frans2025one}, we sample timestep pairs $(s,t)$ by first selecting a discretization level $d$ of the interval $[0,1]$. Each level $d$ is drawn from the set ${0,1,\ldots,d_{\max}}$, where $d_{\max} \geq 1$ is a hyperparameter. This induces a grid of $2^d + 1$ uniformly spaced timesteps on $[0,1]$, allowing up to $K_d = 2^d$ inference steps. In this formulation, $d = 0$ corresponds to one-step inference, $d = 1$ to two-step inference, and so on. To prevent the self-distillation loss from overwhelming the FM loss, we adopt a non-uniform allocation between FM and SD training samples by fixing the probability of sampling boundary cases with $s = t$. Following prior work, we set this probability to $75\%$ in all experiments, resulting in $75\%$ FM samples and $25\%$ SD samples. Details on how this scheme is adapted to each SD setting are provided in \Cref{sec:implem_details}.
\vspace{-0.2cm}

\paragraph{FM training warm-start.} Finally, to prevent the self-distillation objective from dominating early training dynamics, we initialize each Flow Map model by training \emph{without} the SD component in the FM-SD loss for a predefined number of iterations. In our experiments, we first train our model for 5,000 gradient steps using the FM objective, followed by an additional 5,000 gradient steps using the FM-SD loss. Further implementation details are provided in \Cref{sec:implem_details}.

\subsection{Enhancing FlowMapSR via positive-negative prompting guidance } \label{subsec:pos_neg_flow_map}

\begin{figure}[h!]
    \centering
    \vspace{-1.3cm}
    \includegraphics[width=\linewidth]{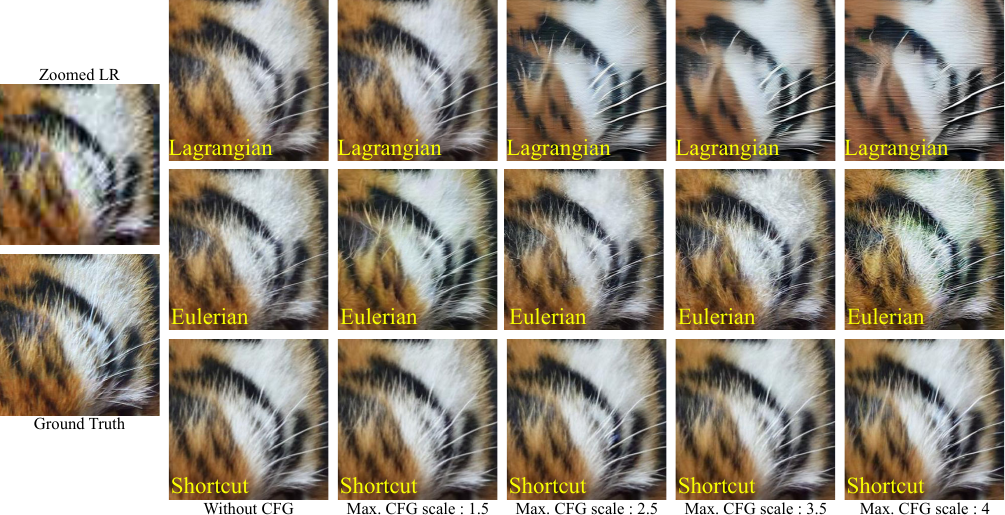}
    \vspace{-0.6cm}
    \caption{\textbf{Qualitative comparison of positive-negative CFG enhancement applied to LSD-, ESD-, and SSD-based FlowMapSR for $\times 4$ upscaling.} The HR reference image exhibits sharp edges and rich high-frequency details. The base FlowMapSR model (``Without CFG'') successfully recovers structural details but lacks perceptual sharpness. Applying positive–negative CFG improves visual realism in the \emph{Shortcut} formulation (e.g., more natural hair textures), while it introduces noticeable degradations and artefacts in the \emph{Lagrangian} and \emph{Eulerian} formulations.}
    \vspace{-0.1cm}
    \label{fig:cfg_impact_all_sd}
\end{figure}

Our base implementation of FlowMapSR, described in \Cref{subsec:base_flow_map}, is unconditional in the sense that it takes only the LR image as input and predicts the corresponding HR image at inference using the update rule \eqref{eq:update}. While this formulation effectively captures meaningful structures from the LR input, see \Cref{fig:cfg_impact_all_sd}, and renders them in a visually plausible manner, the resulting images often lack sharpness and contrast, and therefore fall short in terms of photorealism. To address this limitation, we propose to incorporate CFG into FlowMapSR by adapting the positive–negative prompting strategy introduced in \cite{zhang2024degradation} to our framework, as detailed below. Specifically, we consider the following two generic text prompts:
\begin{itemize}[wide, labelindent=0pt]
\vspace{-0.2cm}
    \item a negative prompt $\mathbf{c}^{\text{neg}}$ : \textit{``oil painting, cartoon, blur, dirty, messy, low quality, deformation, low resolution, oversmooth''}  
    \item a positive prompt $\mathbf{c}^{\text{pos}}$ : \textit{``a high-resolution, 8K, ultra-realistic image with sharp focus, vibrant colors, and natural lighting''}  
\end{itemize}
The objective is to enhance the performance of FlowMapSR conditioned on $\mathbf{c}^{\text{pos}}$ by explicitly modeling poor visual quality through $\mathbf{c}^{\text{neg}}$, and enforcing a strong separation between the two. This strategy provides an effective guidance signal while avoiding reliance on external label extractors, which may be inaccurate or brittle.

Following the CFG formalism introduced in \Cref{subsection:cfg_flowmap}, we treat $\mathbf{c}^{\text{pos}}$ as the \emph{target} conditioning (replacing $\mathbf{c}$) and $\mathbf{c}^{\text{neg}}$ as the \emph{unconditional} conditioning (replacing $\emptyset$). To explicitly learn the notion of degraded quality, we introduce a key modification to the CFG–FM–SD loss related to the SR setting: when the class condition $\mathbf{c}^{\text{pos}}$ is dropped and $\mathbf{c}^{\text{neg}}$ is used instead, we replace the HR target $x_0$ with a synthetic low-quality version $x_0^{\text{neg}} \neq x_1$. This target is constructed to be less degraded than $x_1$, ensuring that the model does not learn a trivial noise trajectory. Concretely, $x_0^{\text{neg}}$ is obtained by applying the same degradation pipeline used to generate $x_1$, except that the downscaling factor is sampled as $s_{\text{down}}^{\text{neg}} \sim \mathrm{U}(s_{\text{down}}, 1)$, where $s_{\text{down}}$ denotes the downscaling factor associated with $x_1$. This design guides the unconditional (negative) model toward generating low-quality versions of the HR image, while still preserving informative structure relative to the input $x_1$ through the difference in upscaling factors. At inference time, we exclusively use the Flow Map model conditioned on $\mathbf{c}^{\text{pos}}$ to generate HR images via the update rule \eqref{eq:update_cfg}. As a result, each text-conditioned Flow Map model (LSD, ESD, or SSD) is enhanced with CFG so as to improve perceptual quality relative to its unconditional counterpart. In early experiments, however, we observe that this strategy is beneficial only for the \emph{Shortcut} formulation, as illustrated in \Cref{fig:cfg_impact_all_sd}. In contrast, the \emph{Lagrangian} and \emph{Eulerian} formulations exhibit noticeable artefacts when combined with CFG; qualitative examples illustrating this failure are provided in \Cref{sec:additional_results}.

\newpage

We attribute the poor performance of our \emph{Lagrangian} and \emph{Eulerian} CFG-based formulations to unstable training dynamics, likely amplified by the use of JVP operators, which are challenging to handle in practice. This instability is consistent with observations reported in prior unconditional implementations, see \cite[Figure 4]{boffi2025buildconsistencymodellearning}, and could potentially be mitigated through specialized training strategies, such as those proposed in \cite[Section 3.4]{sabour2025alignflowscalingcontinuoustime}. In contrast, out \emph{Shortcut} CFG-based formulation exhibits stable training behavior without requiring additional heuristics, which may explain its superior empirical performance. In the following, we denote by $\mathcal{L}(u^\theta_{s,t},\mathrm{I}_t, \mathbf{c}, w;\psi)$ the CFG-based adaptation of the previously established loss $\mathcal{L}(u^\theta_{s,t},\mathrm{I}_t;\psi)$ given in \eqref{eq:loss_unconditional} with positive-negative conditioning $\mathbf{c}$ and guidance scale $w$, where $\mathcal{L}_{\text{FM-SD}}$ is now replaced by $\mathcal{L}_{\text{cfg-FM-SD}}$.

\subsection{Enhancing FlowMapSR via LoRA-based adversarial fine-tuning} \label{subsec:LoRA_flow_map}
Although positive–negative CFG improves the performance of the \emph{Shortcut}-based formulation of FlowMapSR, we still observe residual artefacts in the generated HR images, which are attributable to the guidance mechanism itself (a phenomenon that is also prone to occur in standard image generation tasks). To further mitigate these artefacts and strengthen the alignment between the model outputs and the ground-truth HR distribution, we introduce an additional fine-tuning stage that augments the CFG-based training objective with an adversarial loss, treating the Flow Map model as the generator.

Inspired by recent work \cite{sabour2025alignflowscalingcontinuoustime}, we adopt the Relativistic Pairing GAN (RPGAN) framework \citep{jolicoeur2018relativistic}, which has been shown to be more robust to mode collapse than standard GAN formulations \citep{sun2020towards}. The adversarial training is performed in latent space. Given a discriminator $\mathbf{D}^\phi$ parameterized by a neural network (described in \Cref{sec:implem_details}), we define the following adversarial objective, with separate generator (G-loss) and discriminator (D-loss) terms:
\begin{align*}
    \mathcal{L}^{\text{adv}}(u^\theta_{s,t},\mathrm{I}_t, \mathbf{c},w;\psi,\phi) =\left\{
    \begin{array}{ll}
        \operatorname{Softplus}\left(\mathbf{D}^\phi(\hat{Z}^\text{adv}_0) - \mathbf{D}^\phi(Z_0)\right) + \lambda^{\text{adv}}\mathcal{L}(u^\theta_{s,t},\mathrm{I}_t,\mathbf{c},w;\psi)  & \mbox{ (G-loss)} \\
        \operatorname{Softplus}\left(\mathbf{D}^\phi(Z_0)-\mathbf{D}^\phi(\operatorname{sg}(\hat{Z}^\text{adv}_0))\right) & \mbox{ (D-loss)},
    \end{array}
\right.
\end{align*}
where $\lambda^{\text{adv}} > 0$ controls the contribution of the original loss, $Z_0$ denotes the latent encoding of $X_0$, and $\hat{Z}_0^{\text{adv}}$ is the latent $x_0$-prediction produced by the FlowMapSR model from $X_1$. To fully exploit the expressivity of the Flow Map model, we compute $\hat{Z}_0^{\text{adv}}$ using a two-step prediction (rather than single-step), following \eqref{eq:update_cfg}. In particular, the intermediate state at $t = 1/2$ is detached to avoid the computation of higher-order derivatives, in line with \cite{shen2025adttuningdiffusionmodels}. The parameter $\lambda^{\text{adv}}$ balances adversarial alignment with preservation of the pre-trained FlowMapSR behavior, as excessive adversarial pressure may induce divergence from the original model. As in standard GAN settings, we consider reaching adversarial convergence when we have equality between the “fake” score $\mathbf{D}^\phi(\hat{Z}_0^{\text{adv}})$ and the “real” score $\mathbf{D}^\phi(Z_0)$.

\newpage

\begin{table}[h!]
\caption{\textbf{Comparison of inference efficiency between FlowMapSR and competing SR methods.} We report the average inference time measured on LR RealSR inputs for $\times 4$ upscaling (input resolution $128 \times 128$) using a single H100 GPU. For FlowMapSR, we evaluate three variants with 1, 2, and 4 inference steps. FlowMapSR-1 achieves inference time comparable to OSEDiff-1, while FlowMapSR-2 and FlowMapSR-4 remain competitive with S3Diff-1.}
\label{table:inference}
\begin{center}
\begin{tabular}{l|cccccc}
\toprule
Method    &Real-ESRGAN & OSEDiff-1 & S3Diff-1 & \textbf{FlowMapSR-1}& \textbf{FlowMapSR-2} & \textbf{FlowMapSR-4} \\
\midrule
Inference time (s)   & 0.05 & 0.14 & 0.40 & 0.14 & 0.22 & 0.40\\
\bottomrule
\end{tabular}
\vspace{-0.1cm}
\end{center}
\end{table}

 Owing to the relative nature of RPGAN, which promotes stable training dynamics, we alternate optimization of the G-loss and D-loss at each training step. We found it unnecessary to include additional $R_1$ or $R_2$ regularization terms in the discriminator loss; however, pretraining the discriminator for a small number of iterations was observed to accelerate convergence, and we therefore adopt this strategy in all experiments.

Finally, to keep the fine-tuning stage lightweight while preserving the knowledge acquired by the CFG-based Flow Map model, we restrict adversarial training to Low-Rank Adaptation (LoRA) layers \citep{hu2022lora} within the generator, following the strategy from \cite{chadebec2025flash}. This choice reduces computational cost during training and enables precise control of the LoRA scaling factor at inference time, allowing us to modulate the impact of the adversarial refinement.
\vspace{-0.1cm}
\section{Experiments} \label{sec:expes}

In this section, we evaluate FlowMapSR on synthetic and real-world LR inputs for $\times 4$ and $\times 8$ super-resolution. We first describe the experimental setup in \Cref{subsec:setup}, then present a quantitative comparison with competing SR methods in \Cref{subsec:results}, complemented by representative qualitative examples (with additional results in \Cref{sec:additional_results}, including $\times 2$ super-resolution). These visual comparisons emphasize the recovery of lifelike textures (e.g., fur, human skin, metals) and photorealistic attributes (e.g., depth of field, contrast, color coherence). Finally, in \Cref{subsec:ablation}, we study the key hyperparameters of FlowMapSR that control the trade-off between reconstruction faithfulness and perceptual quality, including the number of inference steps, the LoRA scaling factor at inference, and the maximum positive-negative CFG scale.
\vspace{-0.1cm}
\subsection{Experimental setup} \label{subsec:setup}

\paragraph{FlowMapSR framework.} Throughout this section, our default Flow Map model corresponds to the \emph{Shortcut} variant of FlowMapSR, augmented with positive–negative CFG (\Cref{subsec:pos_neg_flow_map}) and adversarial fine-tuning (\Cref{subsec:LoRA_flow_map}). We do not include results for the \emph{Eulerian} and \emph{Lagrangian} variants, as they consistently underperform on the SR task compared to the \emph{Shortcut} formulation (see \Cref{fig:cfg_impact_all_sd}). Unless stated otherwise, all reported results are obtained by using the guidance strength $w_{\max}=3.5$, setting the LoRA scale as 1.5 and applying 2 inference steps (denoted by \emph{FlowMapSR-2}). The model adopts a UNet architecture following SDXL \citep{podell2024sdxl}, initialized from the image restoration weights of \cite{chadebec2025lbm}. Training is conducted on a single H100 GPU using HR targets of up to $1024 \times 1024$ resolution; additional training details and hyperparameter settings are provided in \Cref{sec:implem_details}. Notably, the same model is used for both $\times 4$ and $\times 8$ upscaling, without any scale-specific conditioning. At inference time, LR inputs are first resized in pixel space using bicubic interpolation according to the target scale, then processed in latent space by the Flow Map model, and finally decoded to obtain the HR output. When memory constraints arise due to large target resolutions, inference is performed after decoding on pixel-space tiles, followed by Gaussian blending.

\paragraph{Testing details.} For synthetic evaluation with available ground truth, we generate LR–HR pairs from the DIV2K validation (DIV2K-Val) set \citep{agustsson2017ntire}, RealSR \citep{cai2019toward}, and DRealSR \citep{wei2020component}, using the degradation pipeline described in \Cref{subsec:base_flow_map}. We recall that this pipeline is a lighter variant of the one proposed in \cite{wang2021realesrgan}, with reduced additional noise, see \Cref{sec:implem_details} for more details. Performance on synthetic datasets is assessed using both reference metrics—PSNR, SSIM \citep{wang2004image}, LPIPS \citep{zhang2018unreasonable}, DISTS \citep{ding2020image}, and FID \citep{heusel2017gans}—which measure fidelity to the HR ground truth, and non-reference metrics—NIQE \citep{zhang2015feature}, MANIQA \citep{yang2022maniqa}, MUSIQ \citep{ke2021musiq}, and CLIPIQA \citep{wang2023exploring}—which better reflect perceptual quality. We additionally present extensive qualitative results on DIV2K-Val and RealSR. For real-world evaluation, we use the RealSet65 dataset \citep{yue2023resshift} and report qualitative comparisons. FlowMapSR is compared against the GAN-based method Real-ESRGAN \citep{wang2021realesrgan} and state-of-the-art single-step diffusion-based methods OSEDiff \citep{wu2024one} and S3Diff \citep{zhang2024degradation} (respectively denoted OSEDiff-1 and S3Diff-1 to emphasize their single-step nature). For fairness, we rely on publicly available code and pre-trained checkpoints to generate HR outputs on the same test sets. While FlowMapSR differs substantially from prior distillation-based approaches in training budget (under 100M parameters for OSEDiff and S3Diff versus over 2B for FlowMapSR), its inference time remains competitive for the default FlowMapSR-2 model, as reported in \Cref{table:inference}.

\newpage

\begin{table}[h!]
\vspace{-1cm}
\caption{ \textbf{Quantitative comparison between FlowMapSR and competing SR methods ($\times 4$ and $\times 8$ upscaling).} The best and second best results are highlighted in \textbf{bold} and \underline{underlined}, respectively.}
\vspace{-0.3cm}
\label{table:upscaling_4_8}
\begin{center}
\resizebox{\textwidth}{!}{%
\begin{tabular}{r|l|ccccc|cccc}
\toprule
\multirow{2}{*}{Dataset ($s_{\text{up}}$)} & \multirow{2}{*}{Method} & \multicolumn{5}{c|}{\underline{Reference metrics}} & \multicolumn{4}{c}{\underline{Non-reference metrics}}\\[0.4em]
 & & PSNR$\uparrow$ & SSIM$\uparrow$ & LPIPS$\downarrow$ & DISTS$\downarrow$ & FID$\downarrow$ & NIQE$\downarrow$ & MANIQA$\uparrow$ & MUSIQ$\uparrow$ & CLIPIQA$\uparrow$\\
\midrule
\multirow{4}{*}{DIV2K-Val ($\times 4$)} & Real-ESRGAN & \bf 23.03 & \bf 0.6880 & 0.2739 & 0.1276 & 21.97 & 3.758 & 0.3536 & 0.6112 & 0.4960 \\
 & OSEDiff-1 & 21.89& \underline{0.6500} & 0.2625 & 0.1107 & 19.23 & 3.373 & \bf 0.4363 & \underline{0.6777} & \bf 0.5974 \\
  & S3Diff-1 & 20.87 & 0.6223 & \bf 0.2475 & \underline{0.1093} & \underline{15.39} & \underline{3.198} & \underline{0.4223} & \bf 0.6780 & \underline{0.5893}\\
  & \cellred{FlowMapSR-2} & \cellred{\underline{22.12}} & \cellred{0.6412} & \cellred{\underline{0.2554}} & \cellred{\bf 0.0995} & \cellred{\bf 13.05} & \cellred{\bf 3.036} & \cellred{0.4010} & \cellred{0.6431} & \cellred{0.5419}\\
\midrule
\multirow{4}{*}{RealSR ($\times 4$)} & Real-ESRGAN & \bf 22.28 & \bf 0.7047& \underline{0.2743}& 0.1939& 126.70& 5.770& 0.4253&0.6497 & 0.4575\\
 & OSEDiff-1 & 21.15& 0.6610& 0.2752& \underline{0.1907}& 106.10& \bf 5.180& 0.4936& 0.6925&0.5288\\
  & S3Diff-1 &\underline{21.75} & \underline{0.6762} &\bf 0.2585 & \bf 0.1775& \bf 94.32&5.661 & \underline{0.5075}&\bf 0.6986 &\underline{0.5511}\\
  & \cellred{FlowMapSR-2} & \cellred{20.42} &\cellred{0.5965}& \cellred{0.3011}& \cellred{0.1964} & \cellred{\underline{95.16}} & \cellred{\underline{5.322}} & \cellred{\bf 0.5362}& \cellred{\underline{0.6975}} & \cellred{\bf 0.5901} \\
  \midrule
\multirow{4}{*}{DrealSR ($\times 4$)} & Real-ESRGAN & \bf 25.22& \bf 0.7771&\bf 0.2700 & \underline{0.1875}&140.31 & 6.954& 0.3876& 0.6001&0.4581\\
 & OSEDiff-1 & 23.54& 0.7173&\underline{0.2884} & 0.1894&114.69 &\underline{5.745} &0.4758 & 0.6565 & 0.5634\\
  & S3Diff-1 & \underline{24.36} & \underline{0.7200}& 0.2886& \bf 0.1818& \bf 105.48& \bf 5.734& \underline{0.4862}& \bf 0.6720&\underline{0.5771}\\
  & \cellred{FlowMapSR-2} & \cellred{23.15}& \cellred{0.6688}& \cellred{0.3056}& \cellred{0.1937}& \cellred{\underline{108.74}}& \cellred{5.779}& \cellred{\bf 0.4863}& \cellred{\underline{0.6600}}&\cellred{\bf 0.5894}\\[0.2em]
\hline \hline\\[-0.8em]
\multirow{4}{*}{DIV2K-Val ($\times 8$)} & Real-ESRGAN & \bf 20.55& \bf 0.6061& 0.3940& 0.1882& \underline{40.98}& 5.209& 0.2505& 0.4612& 0.3317\\
 & OSEDiff-1 & 19.68& 0.5567&0.3909 & \underline{0.1667}&43.41 & 3.394& \bf 0.4080& \bf 0.6511&\bf 0.5759\\
  & S3Diff-1 & 19.30 & 0.5386& \bf 0.3743 & 0.1762& 47.49&\bf 3.069 &0.3468 &\underline{0.6227} &\underline{0.5712}\\
  & \cellred{FlowMapSR-2} &\cellred{\underline{19.90}} & \cellred{\underline{0.5671}}& \cellred{\underline{0.3767}}&\cellred{\bf 0.1469} &\cellred{\bf 28.25}&\cellred{\underline{3.197}} & \cellred{\underline{0.3793}}&\cellred{0.6135} &\cellred{0.5008}\\
\midrule
\multirow{4}{*}{RealSR ($\times 8$)} & Real-ESRGAN & \bf 19.72& \bf 0.6014& \bf 0.3536& \bf 0.2327& 179.34& 6.294& 0.2594&0.4787 &0.2865\\
 & OSEDiff-1 & 18.86& \underline{0.5294}& 0.4174& 0.2526& 180.05& \underline{4.592}& \underline{0.4396}& 0.6454&0.5011\\
  & S3Diff-1 & \underline{18.93}&0.5224 &\underline{0.3828} &\underline{0.2335} & \underline{172.07}&\bf 4.208 & 0.4298& \underline{0.6546}&\underline{0.5217}\\
  & \cellred{FlowMapSR-2} &\cellred{17.96} &\cellred{0.4878} &\cellred{0.4190} &\cellred{0.2576} &\cellred{\bf 163.35} & \cellred{5.173}& \cellred{\bf 0.5042}& \bf \cellred{0.6773}&\cellred{\bf 0.5746}\\
  \midrule
\multirow{4}{*}{DrealSR ($\times 8$)} & Real-ESRGAN & \bf 22.46&\bf 0.6898 &\bf 0.3378&\bf 0.2244&\underline{178.19}&7.246&0.2612&0.4630&0.3252\\
 & OSEDiff-1 & 21.11& 0.5989& 0.4190& 0.2520& 180.56& \underline{4.763}& \underline{0.4216}& 0.6153&0.5201\\
  & S3Diff-1 &\underline{21.17} &0.5631 &0.4150 & \underline{0.2405}& 185.93& \bf 4.390& 0.4099& \underline{0.6321}&\underline{0.5359} \\
  & \cellred{FlowMapSR-2} & \cellred{20.70}& \cellred{\underline{0.5858}}& \cellred{\underline{0.3944}}& \cellred{0.2481}& \cellred{\bf 170.81}& \cellred{5.743}& \cellred{\bf 0.4472}&\cellred{\bf 0.6365} & \cellred{\bf 0.5483}\\
\bottomrule
\end{tabular}
}
\end{center}
\vspace{-0.4cm}
\end{table}

\subsection{Main results} \label{subsec:results}

\begin{figure}[h!]
\vspace{-1.4cm}
    \centering
    \includegraphics[width=\linewidth]{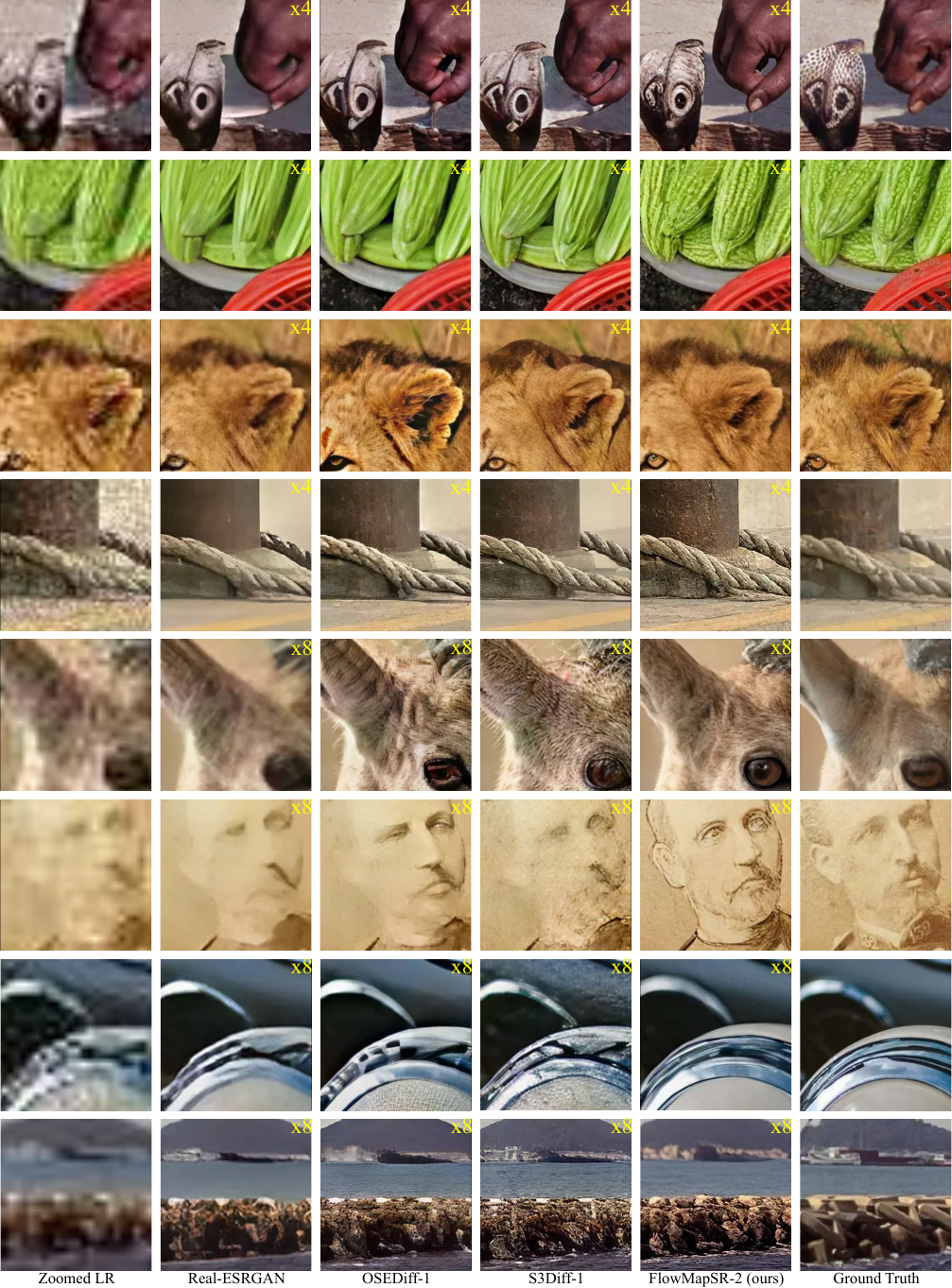}
    \vspace{-0.7cm}
    \caption{\textbf{Qualitative comparison between FlowMapSR and competing SR methods for $\times 4$ (rows 1–4) and $\times 8$ (rows 5–8) upscaling.} Rows 1–3 and 5–7 show examples from DIV2K-Val, while rows 4 and 8 correspond to RealSR samples.}
    \vspace{-2cm}
    \label{fig:x4_x8_upscaling}
\end{figure}

\begin{figure}[h!]
    \vspace{-0.4cm}
    \centering
    \includegraphics[width=\linewidth]{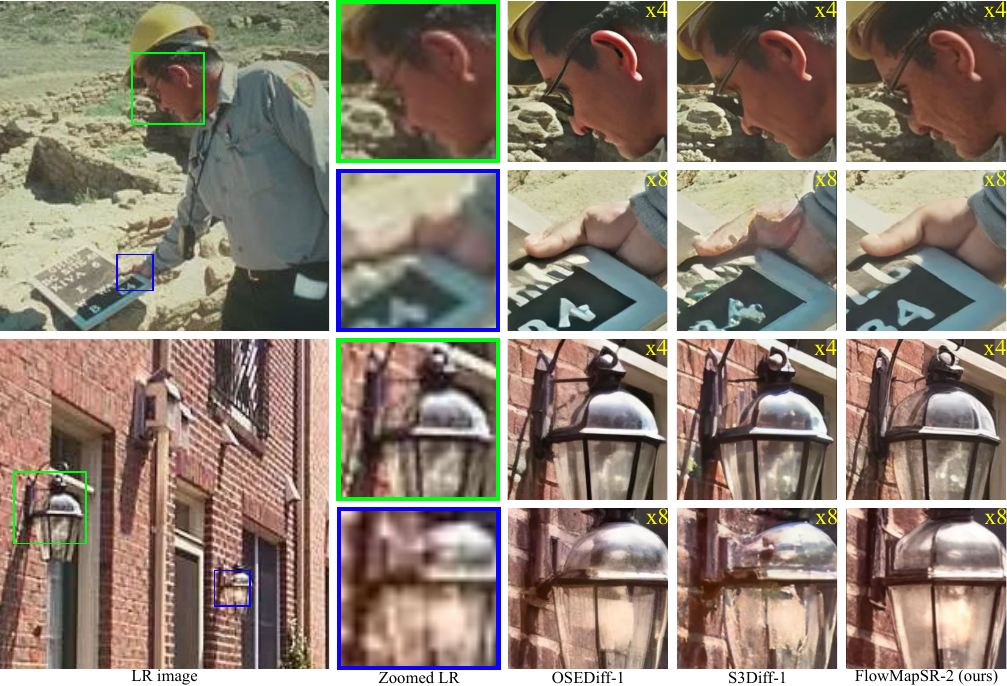}
    \vspace{-0.5cm}
    \caption{\textbf{Qualitative comparison between FlowMapSR and competing diffusion-based SR methods on real-world LR inputs for $\times 4$ and $\times 8$ upscaling.} The two LR images are taken from the RealSet65 dataset. For each image, a \tcmv{green} region is selected for $\times 4$ upscaling and a \tcmb{blue} region for $\times 8$ upscaling. FlowMapSR-2 consistently achieves the best balance between reconstruction faithfulness and photorealism.}
    \label{fig:real_main}
\end{figure}

We report quantitative results for $\times 4$ and $\times 8$ upscaling in \Cref{table:upscaling_4_8}, from which we draw the following observations.
\begin{enumerate}[wide, labelindent=0pt, label=(\alph*)]
    \item On DIV2K-Val, FlowMapSR achieves strong performance on reference-based metrics, particularly perceptual measures such as LPIPS, DISTS, and FID, compared to competing methods. This high fidelity to the DIV2K-Val ground-truth images is consistent with the qualitative results shown in \Cref{fig:x4_x8_upscaling}, see respectively the rows 1-3 for $\times 4$ upscaling and the rows 4-7 for $\times 8$ upscaling. In particular, FlowMapSR exhibits faithful recovery of textures and details (e.g., human skin and attributes, animal fur), geometric structures, lighting, environmental context, and depth of field. While these qualitative examples demonstrate strong photorealism—often surpassing competing methods—this advantage is not always reflected in non-reference metrics, where OSEDiff-1 tends to achieve higher scores. Visually, this may be explained by the stronger sharpness and contrast produced by OSEDiff-1, which benefit such metrics but come at the expense of reference fidelity. We attribute the slightly lower non-reference scores of FlowMapSR in this setting to occasional boundary artefacts introduced by Gaussian tiling, see \Cref{sec:additional_results}, which is not required for processing RealSR and DRealSR images. Improving this tiling strategy is left for future work.
    \item On RealSR and DRealSR, the trend is reversed: FlowMapSR achieves the best average performance on non-reference metrics for $\times 4$ and $\times 8$ upscaling on both datasets, while still maintaining competitive DISTS and FID scores, especially at higher scaling factor. Although Real-ESRGAN attains strong reference-based scores on these datasets, this should be interpreted with caution. Visual inspection in \Cref{fig:x4_x8_upscaling} (row 4 for $\times 4$, and row 8 for $\times 8$) reveals that Real-ESRGAN often produces visibly low-quality outputs. Despite being counter-intuitive, high reference-based scores and poor visual quality are not contradictory here; both reflect the limited visual quality of the RealSR and DrealSR ground truth, thereby revealing the limitations of reference metrics in this case. 
    In contrast, while FlowMapSR may appear less faithful based on reference metrics, it consistently produces higher-quality images, with more realistic textures and convincing depth of field areas, where S3Diff comparatively struggles.
\end{enumerate}
Finally, we present qualitative comparisons on real-world inputs from RealSet65 in \Cref{fig:real_main}, where we show SR results for selected regions of the LR images. In this fully real-world setting, FlowMapSR consistently produces the most photorealistic and faithful outputs for both $\times 4$ and $\times 8$ upscaling, particularly in terms of texture realism and depth of field effects. Overall, FlowMapSR demonstrates strong and consistent performance across both synthetic and real-world scenarios, using a single model without any upscaling-specific conditioning. While this behavior is only partially captured by quantitative metrics, it is clearly reflected in qualitative results. Additional comparisons on DIV2K-Val, RealSR, and RealSet65 for both $\times 4$ and $\times 8$ upscaling are provided in \Cref{sec:additional_results}.

\newpage

\newpage

\subsection{Ablation studies} \label{subsec:ablation}
We analyze the influence of three key hyperparameters of FlowMapSR: the number of inference steps, the LoRA scaling factor applied at inference, and the maximum CFG scale used during training. For each setting, we report quantitative results on DIV2K-Val and RealSR for $\times 4$ and $\times 8$ upscaling, complemented by qualitative comparisons for the first two ablations. Additional visual results are provided in \Cref{sec:additional_results}.

\begin{figure}[h!]
    \centering
    \includegraphics[width=\linewidth]{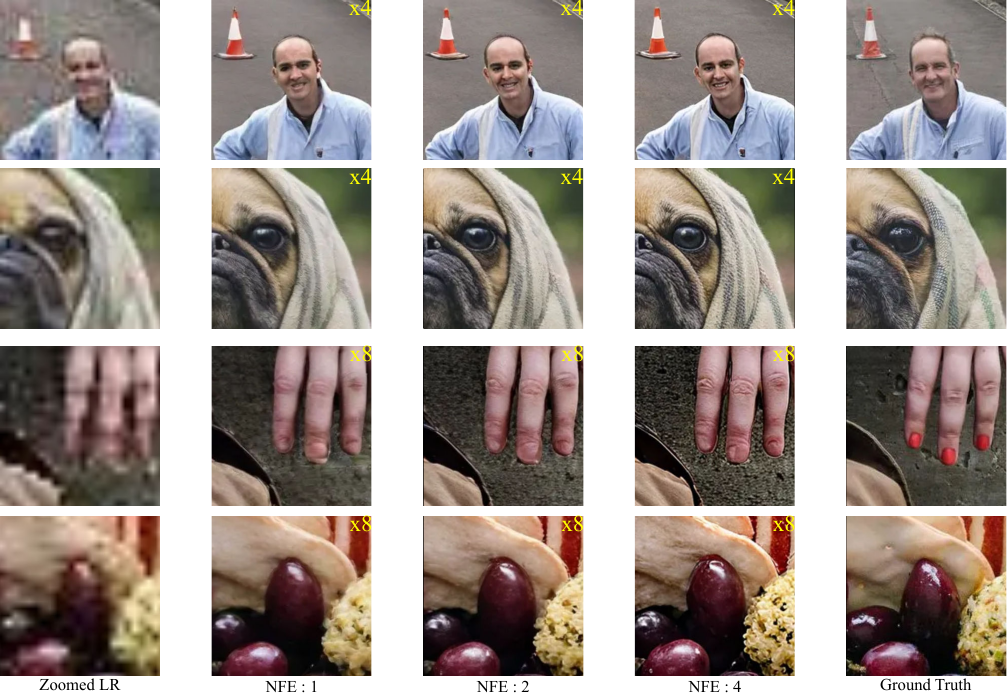}
    \vspace{-0.5cm}
    \caption{\textbf{Qualitative comparison of FlowMapSR with varying numbers of inference steps (NFE) for $\times 4$ (rows 1–2) and $\times 8$ (rows 3–4) upscaling.}
The LR inputs are drawn from the DIV2K-Val dataset. Increasing NFE improves perceptual quality, while fewer steps yield higher faithfulness to the reference, illustrating the trade-off quantified in \Cref{table:ablation_num_steps}.}
    \label{fig:ablation_num_steps}
\end{figure}

\paragraph{Impact of the number of inference steps.}
Our Flow Map formulation allows the number of inference steps (NFE) to be chosen as a power of two; we evaluate 1, 2, and 4 steps. As shown in \Cref{table:ablation_num_steps}, using a single step consistently yields higher fidelity to the reference images, while increasing NFE improves model expressivity and perceptual quality, as reflected by non-reference metrics. These results highlight a clear trade-off: increasing NFE is not always beneficial. Qualitative examples in \Cref{fig:ablation_num_steps} illustrate that larger NFE can degrade depth-of-field consistency (rows 1 and 3) and introduce overly shiny or glassy textures (rows 2 and 4), whereas a single step, while more faithful, lacks visual refinement. Overall, using two inference steps, as in our default configuration, provides an effective balance between faithfulness and perceptual quality.

\begin{figure}[h!]
%\vspace{-0.5cm}
    \centering
    \includegraphics[width=\linewidth]{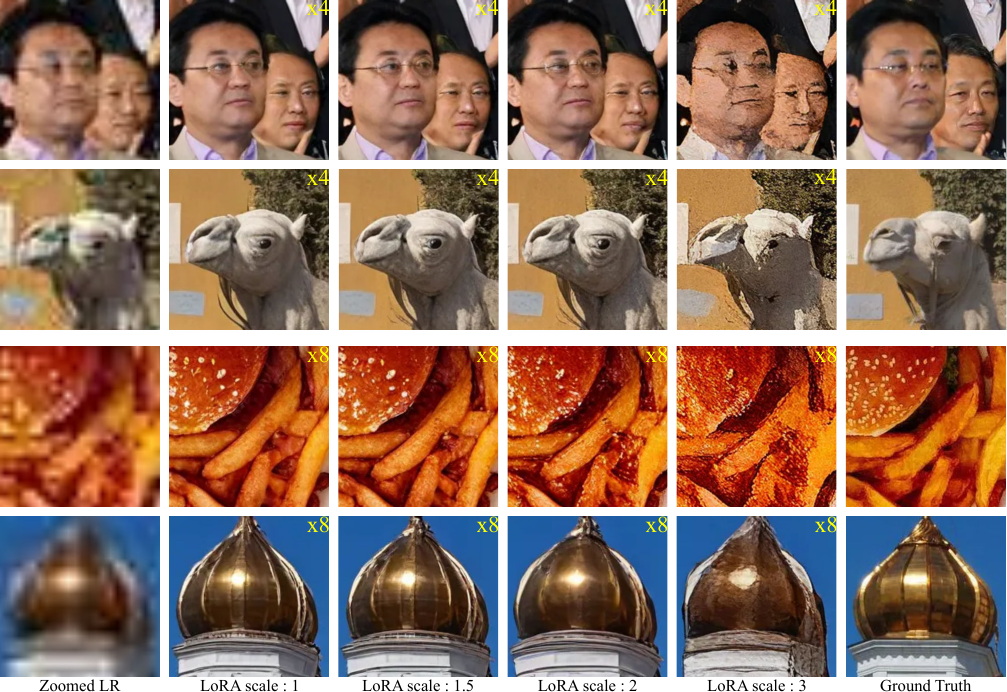}
    \vspace{-0.5cm}
    \caption{\textbf{Qualitative comparison of FlowMapSR with varying LoRA scale for $\times 4$ (rows 1–2) and $\times 8$ (rows 3–4) upscaling.} Inputs are from the DIV2K-Val dataset. Smaller LoRA scales better preserve fidelity to the reference, while larger scales favor sharpness (eventually degrading quality with LoRA scale set to 3), consistent with the trade-off in \Cref{table:ablation_lora_scale}.}
    \vspace{-1cm}
    \label{fig:ablation_lora_scale}
\end{figure}

\paragraph{Impact of the LoRA scale at inference.}
A similar trade-off is observed with respect to the LoRA scaling factor, which controls the influence of the adversarially fine-tuned parameters at inference. Increasing the LoRA scale (e.g., from 1 to 2) enhances sharpness and contrast, as illustrated in \Cref{fig:ablation_lora_scale}. However, this comes at the cost of reduced reconstruction faithfulness and, in some cases, degraded photorealism, particularly in depth-of-field rendering (see rows 1 and 2). This effect becomes more pronounced for larger values (e.g., LoRA scale 3), which lead to clearly suboptimal results. The quantitative results in \Cref{table:ablation_lora_scale} confirm this behavior, with higher reference-based scores at lower LoRA scales and improved non-reference metrics at larger scales. In practice, the default setting with a LoRA scale of 1.5 offers a good compromise between fidelity and perceptual enhancement.

\newpage

\begin{table}[t]
\vspace{-0.9cm}
\caption{\textbf{Quantitative comparison of FlowMapSR with varying number of inference steps (NFE) ($\times 4$ and $\times 8$ upscaling).} All FlowMapSR variants use $w_{\max}=3.5$ and a LoRA scale of $1.5$, consistent with the FlowMapSR-2 configuration. The best and second best results are highlighted in \textbf{bold} and \underline{underlined}, respectively. \tcmrr{Red} cells denote the default FlowMapSR-2 configuration used in \Cref{table:upscaling_4_8} and \Cref{fig:x4_x8_upscaling}, while \tcmbb{blue} cells mark the best results when also considering competing methods reported in \Cref{table:upscaling_4_8}.}
\vspace{-0.3cm}
\label{table:ablation_num_steps}
\begin{center}
\resizebox{\textwidth}{!}{%
\begin{tabular}{r|c|ccccc|cccc}
\toprule
\multirow{2}{*}{Dataset ($s_{\text{up}}$)} & \multirow{2}{*}{NFE} & \multicolumn{5}{c|}{\underline{Reference metrics}} & \multicolumn{4}{c}{\underline{Non-reference metrics}}\\[0.4em]
 & & PSNR$\uparrow$ & SSIM$\uparrow$ & LPIPS$\downarrow$ & DISTS$\downarrow$ & FID$\downarrow$ & NIQE$\downarrow$ & MANIQA$\uparrow$ & MUSIQ$\uparrow$ & CLIPIQA$\uparrow$\\
\midrule
\multirow{3}{*}{DIV2K-Val (x4)} &  1 & \bf 22.89 & \bf 0.6703& \bf \cellblue{0.2464}& \bf \cellblue{0.0986}&14.03 & 3.750& 0.3810&0.6248 &0.4995\\
  & \cellred{2}&\cellred{\underline{22.12}} & \cellred{\underline{0.6412}} & \cellred{\underline{0.2554}} & \cellred{\underline{0.0995}} & \bf \cellred{13.05} & \cellred{\underline{3.036}} & \cellred{\underline{0.4010}} & \cellred{\underline{0.6431}} & \cellred{\underline{0.5419}
} \\
  & 4 & 20.94& 0.5881& 0.3033& 0.1126& \underline{13.55}& \bf \cellblue{2.775}&\bf 0.4134 & \bf 0.6575&\bf 0.5848\\
\midrule
\multirow{3}{*}{RealSR (x4)} & 1 & \bf 21.74 & \bf 0.6517& \bf 0.2636& \bf 0.1790& \bf \cellblue{90.54}& 5.630&0.4949 &0.6764 &0.5196\\
  &  \cellred{2}& \cellred{\underline{20.42}} &\cellred{\underline{0.5965}}& \cellred{\underline{0.3011}}& \cellred{\underline{0.1964}} & \cellred{\underline{95.16}} & \cellred{\bf 5.322} & \cellred{\underline{0.5362}}& \cellred{\underline{0.6975}} & \cellred{\underline{0.5901}}\\
  &  4 & 19.15 & 0.5322& 0.3598& 0.2174&103.86 & \underline{5.342}&\bf  \cellblue{0.5607}& \bf \cellblue{0.7095}&\bf \cellblue{0.6204}\\[0.2em]
\hline \hline\\[-0.8em]
\multirow{3}{*}{DIV2K-Val (x8)} &  1 &  \bf \cellblue{20.75}& \bf 0.5981& \bf \cellblue{0.3568}&\underline{0.1499} &\bf \cellblue{27.97} & 3.857&0.3537 & 0.5914&0.4648\\
  &  \cellred{2}& \cellred{\underline{19.90}} & \cellred{\underline{0.5671}}& \cellred{\underline{0.3767}}&\cellred{\bf 0.1469} &\cellred{\underline{28.25}}&\cellred{\underline{3.197}} & \cellred{\underline{0.3793}}&\cellred{\underline{0.6135}} &\cellred{\underline{0.5008}} \\
  & 4 & 18.85 &0.5223 & 0.4249&0.1525 &29.82 &\bf \cellblue{3.077} &\bf 0.3987 & \bf 0.6311&\bf 0.5449\\
\midrule
\multirow{3}{*}{RealSR (x8)} & 1 &\bf 19.28  & \bf 0.5486& \bf 0.3886&\bf 0.2458 & 166.73& 5.758&0.4536 &0.6495 &0.5178\\
  & \cellred{2}& \cellred{\underline{17.96}} &\cellred{\underline{0.4878}} &\cellred{\underline{0.4190}} &\cellred{\underline{0.2576}} &\cellred{\bf 163.35} & \cellred{\bf 5.173}& \cellred{\underline{0.5042}}& \cellred{\underline{0.6773}}&\cellred{\underline{0.5746}} \\
  &  4 & 16.67 & 0.4179& 0.4773&0.2755 &\underline{164.99} & \underline{5.535}&\bf \cellblue{0.5309} &\bf \cellblue{0.6970} &\bf \cellblue{0.6127}\\
  \bottomrule
\end{tabular}
}
\end{center}
\end{table}

\newpage

\begin{table}[h!]
\vspace{-0.9cm}
\caption{\textbf{Quantitative comparison of FlowMapSR with varying LoRA scale ($\times 4$ and $\times 8$ upscaling).} All FlowMapSR variants use $w_{\max}=3.5$ with 2 inference steps, consistent with the FlowMapSR-2 configuration. The best and second best results are highlighted in \textbf{bold} and \underline{underline}, respectively. \tcmrr{Red} cells correspond to the default version of FlowMapSR-2 used in \Cref{table:upscaling_4_8} and \Cref{fig:x4_x8_upscaling}.} 
\vspace{-0.3cm}
\label{table:ablation_lora_scale}
\begin{center}
\resizebox{\textwidth}{!}{%
\begin{tabular}{r|c|ccccc|cccc}
\toprule
\multirow{2}{*}{Dataset ($s_{\text{up}}$)} & \multirow{2}{*}{\shortstack{LoRA\\scale}} & \multicolumn{5}{c|}{\underline{Reference metrics}} & \multicolumn{4}{c}{\underline{Non-reference metrics}}\\[0.4em]
 & & PSNR$\uparrow$ & SSIM$\uparrow$ & LPIPS$\downarrow$ & DISTS$\downarrow$ & FID$\downarrow$ & NIQE$\downarrow$ & MANIQA$\uparrow$ & MUSIQ$\uparrow$ & CLIPIQA$\uparrow$\\
\midrule
\multirow{4}{*}{DIV2K-Val (x4)} &  1 & \bf 22.29 & \bf 0.6520& \underline{0.2571}& \bf 0.0958& \bf 12.92&3.170& 0.3865&0.6309& 0.5138\\
 & \cellred{1.5} & \cellred{\underline{22.12}} & \cellred{\underline{0.6412}} & \cellred{\bf 0.2554} & \cellred{\underline{0.0995}} &  \cellred{\underline{13.05}} & \cellred{\underline{3.036}} & \cellred{0.4010} & \cellred{\underline{0.6431}} & \cellred{0.5419
}\\
  &  2& 21.83& 0.6224& 0.2675&0.1093 & 13.43 &\bf 2.928 & \underline{0.4059}&\bf 0.6456& \underline{0.5649}\\
  & 3 &20.21 &0.5232 &0.3928 & 0.1800& 19.20& 3.569& \bf 0.4149&0.6323& \bf 0.6486\\
\midrule
\multirow{4}{*}{RealSR (x4)} &  1 & \bf 20.50& \bf 0.6084& \bf 0.2891 & \bf 0.1907 &\bf 92.76 & \bf 5.252& \underline{0.5285}&\underline{0.6961}& 0.5753\\
 & \cellred{1.5} & \cellred{\underline{20.42}} &\cellred{\underline{0.5965}}& \cellred{\underline{0.3011}}& \cellred{\underline{0.1964}} & \cellred{\underline{95.16}} & \cellred{\underline{5.322}} & \cellred{\bf 0.5362}& \cellred{\bf 0.6975} & \cellred{\underline{0.5901}}\\
  &  2&20.22 &0.5753 &0.3274 &0.2109 & 103.50& 5.659&0.5271 &0.6860& \bf 0.5967\\
  & 3 & 19.39& 0.5280& 0.4140&0.2526 & 137.74& 7.134&0.4991 &0.6662& 0.6009\\[0.2em]
\hline \hline\\[-0.8em]
\multirow{4}{*}{DIV2K-Val (x8)} &  1 & \bf 20.08& \bf 0.5782& \underline{0.3792}& \bf 0.1415& \bf 27.86& 3.313& 0.3671&0.6021& 0.4721\\
 & \cellred{1.5} & \cellred{\underline{19.90}} & \cellred{\underline{0.5671}}& \cellred{\bf 0.3767}&\cellred{\underline{0.1469}} &\cellred{\underline{28.25}}&\cellred{\underline{3.197}} & \cellred{\underline{0.3793}}&\cellred{\underline{0.6135}} &\cellred{0.5008}\\
  &  2& 19.62& 0.5457& 0.3888&0.1645 & 30.45&\bf 3.022 & 0.3788&0.6115& \underline{0.5367}\\
  & 3 &18.71 &0.4670 & 0.4987&0.2392 &43.63 & 3.592& \bf 0.3938&\bf 0.6136& \bf 0.6600\\
\midrule
\multirow{4}{*}{RealSR (x8)} &  1 & \bf 18.07&\bf 0.5019 &\bf 0.4041 &\bf 0.2480 & \bf 158.81 &\bf 5.163&\underline{0.4989} & \bf 0.6785& 0.5634\\
 & \cellred{1.5} & \cellred{\underline{17.96}} &\cellred{\underline{0.4878}} &\cellred{\underline{0.4190}} &\cellred{\underline{0.2576}} &\cellred{\underline{163.35}} & \cellred{\underline{5.173}}& \cellred{\bf 0.5042}& \cellred{\underline{0.6773}}&\cellred{0.5746}\\
  &  2& 17.80 & 0.4638& 0.4563& 0.2797&182.77 & 5.537& 0.4791&0.6580& \underline{0.5814}\\
  & 3 & 17.76 & 0.4410& 0.5492& 0.3148&229.96 & 6.784& 0.4491&0.6226& \bf 0.5954\\
  \bottomrule
\end{tabular}
}
\end{center}
\end{table}

\newpage
\begin{table}[h!]
\vspace{-1.6cm}
\caption{\textbf{Quantitative comparison of FlowMapSR with maximal positive-negative CFG scale $w_{\max}$ ($\times 4$ and $\times 8$ upscaling).} The best and second best results are highlighted in \textbf{bold} and \underline{underlined}, respectively. \tcmbb{Blue} cells mark the best results when also considering competing methods reported in \Cref{table:upscaling_4_8}. All FlowMapSR variants use 2 inference steps, consistent with the FlowMapSR-2 configuration, but are obtained \textbf{without adversarial fine-tuning}.}
\label{table:ablation_cfg}
\vspace{-0.4cm}
\begin{center}
\resizebox{\textwidth}{!}{%
\begin{tabular}{r|c|ccccc|cccc}
\toprule
\multirow{2}{*}{Dataset ($s_{\text{up}}$)} & \multirow{2}{*}{\shortstack{Max.\\ CFG\\scale}} & \multicolumn{5}{c|}{\underline{Reference metrics}} & \multicolumn{4}{c}{\underline{Non-reference metrics}}\\[0.4em]
 & & PSNR$\uparrow$ & SSIM$\uparrow$ & LPIPS$\downarrow$ & DISTS$\downarrow$ & FID$\downarrow$ & NIQE$\downarrow$ & MANIQA$\uparrow$ & MUSIQ$\uparrow$ & CLIPIQA$\uparrow$\\
\midrule
\multirow{5}{*}{DIV2K-Val (x4)} &  2 & \underline{23.15}& \underline{0.6759}& \underline{0.2568}& 0.1015& \cellblue{\bf 11.93}& 3.489& 0.3484& 0.6042&0.4605\\
 & 2.5 & 22.98&0.6675 & 0.2578& \cellblue{\bf 0.0987}& \underline{12.08}& \underline{3.314}& 0.3584& \underline{0.6068}&0.4794\\
  &  3& \cellblue{\bf23.48}& \cellblue{\bf0.6913}& 0.2658& 0.1052& 12.50& 4.052&0.3344 & 0.5822&0.4415\\
  &  3.5& 22.12& 0.6413& \bf 0.2554& \underline{0.0995}&13.05 & \bf \cellblue{3.036}& \underline{0.4008}& \bf 0.6433&\underline{0.5415}\\
  &  4& 20.37& 0.5530& 0.3602& 0.1612& 15.34& 5.072& \bf 0.4230& \underline{0.6068}&\cellblue{\bf 0.6195}\\
\midrule
\multirow{5}{*}{RealSR (x4)} &  2 & \underline{22.32}& \underline{0.6627}& \underline{0.2402}& 0.1692& 88.92& 4.911& 0.4473& 0.6630&0.4860\\
 & 2.5 & 22.08&0.6545 & 0.2450&\underline{0.1688} & \underline{87.15}& \underline{4.677}& 0.4634& 0.6691&0.5106\\
  &  3& \cellblue{\bf 22.77}& \bf 0.6924& \cellblue{\bf 0.2180}& \cellblue{\bf 0.1582}&\cellblue{\bf 81.01} & \bf 5.278& 0.4435& 0.6602&0.4755\\
  &  3.5& 20.43& 0.5968&0.3009 & 0.1964& 94.71& 5.331& \underline{0.5368}&\bf 0.6976 &\underline{0.5910}\\
  &  4& 18.75& 0.5003&0.4191 & 0.2469&107.96& 8.727& \cellblue{\bf 0.5513}& \underline{0.6777}&\cellblue{\bf 0.6371}\\[0.2em]
\hline \hline\\[-0.8em]
\multirow{5}{*}{DIV2K-Val (x8)} &  2 & 20.27& 0.5807&\bf 0.3634& \bf \cellblue{0.1420}& \cellblue{\bf 27.59}& 3.331& 0.3601& \underline{0.6118}&0.4737\\
 & 2.5 & \underline{20.45}& \underline{0.5818}& 0.3771& 0.1494& 28.57&\underline{3.270} & 0.3582& 0.5926&0.4622\\
  &  3& \cellblue{\bf 20.87}& \cellblue{\bf 0.6091}& \underline{0.3726}& 0.1499& 28.75& 4.158& 0.3237&0.5629 &0.4183\\
  &  3.5& 19.90& 0.5671& 0.3767& \underline{0.1469}& \underline{28.31}& \bf 3.199& \underline{0.3794}& \bf 0.6135&\underline{0.5006}\\
  &  4& 19.53& 0.5334& 0.4639& 0.1988& 30.71& 4.553& \bf 0.3907& 0.5634&\bf 0.5510\\
\midrule
\multirow{5}{*}{RealSR (x8)} &  2 & 18.78& \underline{0.5241}& \underline{0.3695}& \underline{0.2350}& \cellblue{\bf 151.44}& 4.524& 0.4542& 0.6673&0.5238\\
 & 2.5 &\underline{18.98} & 0.5168& 0.3835& 0.2404& 155.66&\bf \underline{4.474} & 0.4540& 0.6628&0.5383\\
  &  3& \bf 19.40& \bf 0.5656 & \cellblue{\bf 0.3465}& \cellblue{\bf 0.2280}& \underline{154.32}& 5.131& 0.4342& \underline{0.6590}&0.4978\\
  &  3.5& 17.96& 0.4877& 0.4188& 0.2570&162.22 & 5.178& \underline{0.5034}& \cellblue{\bf 0.6776}&\underline{0.5749}\\
  &  4& 17.64& 0.4316& 0.5353& 0.2996& 175.57& 8.143& \cellblue{\bf 0.5140}& 0.6408&\cellblue{\bf 0.6261}\\
  \bottomrule
\end{tabular}
}
\end{center}
\end{table}

\vspace{-0.6cm}
\paragraph{Impact of the positive–negative CFG strength.}

Finally, we observe a similar trade-off when studying the impact of the CFG scale used to train FlowMapSR (\textbf{without adversarial fine-tuning}), as shown in \Cref{table:ablation_cfg}. Increasing the maximum guidance strength $w_{\max}$ consistently improves perceptual quality, as indicated by non-reference metrics, with $w_{\max}=4$ yielding CLIPIQA scores that surpass competing methods. This gain, however, comes at the cost of reconstruction faithfulness: reference-based metrics degrade at higher guidance strengths, reflecting weaker alignment with the HR ground truth. Lower CFG scales favor fidelity, though the optimal value is not straightforward: $w_{\max}=2$ performs best in terms of FID and DISTS, while PSNR and SSIM peak at $w_{\max}=3$. Overall, we adopt $w_{\max}=3.5$ as a balanced compromise between perceptual quality and faithfulness. Although adversarial fine-tuning does not change the quantitative metrics for this setting (see \Cref{table:upscaling_4_8}), it leads to clear visual improvements in the generated HR images (see \Cref{fig:all_gan} in \Cref{sec:additional_results}).
\vspace{-0.2cm}

\section{Conclusion}
\vspace{-0.2cm}

In this paper, we introduced FlowMapSR, a diffusion-based framework for image super-resolution built on Flow Map generative models and designed for fast inference. Unlike prior diffusion-based SR approaches based on teacher–student distillation, FlowMapSR directly trains a large, expressive model while maintaining efficiency. The framework supports all three Flow Map formulations (\emph{Lagrangian}, \emph{Eulerian}, and \emph{Shortcut}) and incorporates two SR-specific enhancements: a unified CFG-based positive–negative prompting strategy and a lightweight LoRA-based adversarial fine-tuning stage for controllable perceptual refinement at inference.

Empirically, the \emph{Shortcut} formulation best accommodates these enhancements, while the \emph{Lagrangian} and \emph{Eulerian} variants suffer from unstable training. We conjecture that this limitation could be alleviated through finer-grained loss control \citep{sabour2025alignflowscalingcontinuoustime}, which we leave for future work. Using the \emph{Shortcut} variant as our default configuration, FlowMapSR achieves a strong balance between reconstruction faithfulness and photorealism, outperforming competing diffusion-based SR methods in qualitative realism. It produces more lifelike textures (especially, for human subjects), improved depth-of-field rendering, and fewer overly sharp or glossy artefacts, all using a single model without scale- or degradation-specific conditioning. While training is more expensive than distillation-based approaches, inference time remains competitive. Finally, our ablations studies show that the fundamental SR trade-off is explicitly controlled by FlowMapSR hyperparameters: increasing the number of inference steps, CFG strength, or LoRA scale improves perceptual quality, while lower values favor fidelity. 

A current limitation of our approach is the reliance on Gaussian tiling at very high resolutions, which can introduce mild blurring at the boundaries of generated images; such artefacts could likely be alleviated through more advanced tiling strategies. Additionally, our procedure may exhibit noticeable color shifts, a phenomenon commonly observed in diffusion models \citep{deck2023easing}. In practice, FlowMapSR could be combined with GPU-like post-processing techniques to correct these shifts, as explored for instance in \cite{feydy2019interpolating}.
We hope our results motivate further exploration of Flow Map models for other image-to-image translation tasks, such as object removal and relighting, depth and normal estimation.

\newpage
\bibliographystyle{iclr2026_conference}
\bibliography{main.bib}

\newpage
\appendix

\section*{Organization of the supplementary}

The appendix is organized as follows. \Cref{sec:theory_details} first presents complementary theoretical material related to the Flow Map models introduced in \Cref{sec:flow_map_theory}. \Cref{sec:implem_details} then provides additional implementation details regarding the data, model architecture, training procedure, and inference setup used in our experiments. Finally, \Cref{sec:additional_results} reports further qualitative results for $\times 4$ and $\times 8$ super-resolution, including comparisons with competing methods on additional synthetic and real-world samples, as well as extended analyses of the ablation studies discussed in \Cref{sec:expes}.

\vspace{-0.2cm}
\section{Theoretical details} \label{sec:theory_details}

\subsection{Background on Flow Matching and Flow Map models} \label{subsec:flowmap_general}
In the case where the interpolant $\mathrm{I}_t$ admits the general form $\mathrm{I}_t=\alpha_t X_0 + \beta_t X_1$, a velocity field candidate for the generative ODE \eqref{eq:ODE} is given by $v_t(x)=\mathbb{E}[v_{\text{FM}}^{\text{target}}|\mathrm{I}_t=x]$ where $v_{\text{FM}}^{\text{target}}=\dot{\mathrm{I}}_t=\dot{\alpha}_t X_0 + \dot{\beta}_t X_1$ is the conditional target velocity. Following \cite{boffi2025buildconsistencymodellearning}, the loss target in the SD optimization problem \eqref{eq:sd_loss} related to this general SI writes as follows
\begin{enumerate}
    \item \textit{Lagrangian} SD (LSD) loss
    \begin{align}\label{eq:general_lsd_loss}
        u_{\text{LSD}}^{\text{target}} = u^\theta_{s,s}\left(\mathrm{I}_t -(t-s) u_{s,t}^\theta(\mathrm{I}_t)\right) + (t-s) \partial_s u^\theta_{s,t}(\mathrm{I}_t) \eqsp ,
    \end{align}
    \item \textit{Eulerian} SD (ESD) loss
    \begin{align}\label{eq:general_esd_loss}
        u_{\text{ESD}}^{\text{target}} = u^\theta_{t,t}(\mathrm{I}_t)-(t-s)\left(\nabla u^\theta_{s,t}(\mathrm{I}_t) u^\theta_{t,t}(\mathrm{I}_t) + \partial_t u^\theta_{s,t}(\mathrm{I}_t)\right)\eqsp ,
    \end{align}
    \item \textit{Shortcut} SD (SSD) loss
    \begin{align} \label{eq:general_ssd_loss}
        u_{\text{SSD}}^{\text{target}}= \frac{t-r}{t-s}u^\theta_{r,t}(\mathrm{I}_t) + \frac{r-s}{t-s}u^\theta_{s,r}\left(\mathrm{I}_t -(t-r) u^\theta_{r,t}(\mathrm{I}_t)\right), \quad r \sim q(r|s,t)\eqsp ,
    \end{align}
\end{enumerate}
where $q(r|t,s)$ is a time distribution on $\{r\in [0,1]:s<r<t\}$. These expressions are derived from straightforward computations on the Flow Map \eqref{eq:param_euler}, summarized below as an informal counterpart to the proofs in \cite{boffi2025buildconsistencymodellearning}.

\begin{proof}
\textbf{Lagrangian setting.} By computing the derivative of the Flow Map \eqref{eq:param_euler} evaluated in the trajectory $(X_t)_{t\in [0,T]}$ with respect to time variable $s$, we have
    \begin{align*}
        \frac{\rmd X_{s,t}(X_t)}{\rmd s} &= u_{s,t}(X_t) -(t-s)\partial_s u_{s,t}(X_t)  \eqsp ,
    \end{align*}
By definition of the Flow Map given in \eqref{eq:flowmap_def}, we also have that $\rmd X_{s,t}(X_t)/\rmd s = v_s(X_{s,t}(X_t))$. By combining these two results, we obtain the \textit{Lagrangian characterization}
\begin{align*}
    u_{s,t}(X_t) &= v_s(X_{s,t}(X_t)) + (t-s) \partial_s u_{s,t}(X_t)\\
     &= u_{s,s}(X_{s,t}(X_t)) + (t-s) \partial_s u_{s,t}(X_t) \eqsp .
\end{align*}
After replacing $u$ by $u^\theta$ in the previous equality, we obtain the LSD target \eqref{eq:general_lsd_loss} from the right term.

\noindent\textbf{Eulerian setting.} By computing the derivative of the Flow Map \eqref{eq:param_euler} evaluated in the trajectory $(X_t)_{t\in [0,T]}$ with respect to time variable $t$, we have
    \begin{align*}
        \frac{\rmd X_{s,t}(X_t)}{\rmd t} &= \frac{\rmd X_t}{\rmd t} -u_{s,t}(X_t)-(t-s) \frac{\rmd u_{s,t}(X_t)}{\rmd t} \\
        & = v_t(X_t) -u_{s,t}(X_t)-(t-s)(\nabla u_{s,t}(X_t) v_t(X_t) + \partial_t u_{s,t}(X_t))\eqsp .
    \end{align*}
By definition of the Flow Map given in \eqref{eq:flowmap_def}, we also have that $\rmd X_{s,t}(X_t)/\rmd t = v_t(X_t) -v_t(X_t)=0$. By combining these two results, we obtain the \textit{Eulerian characterization}
    \begin{align*}
        u_{s,t}(X_t) &= v_t(X_t) -(t-s)(\nabla u_{s,t}(X_t) v_t(X_t) + \partial_t u_{s,t}(X_t))\\
        &= u_{t,t}(X_t) -(t-s)(\nabla u_{s,t}(X_t) u_{t,t}(X_t) + \partial_t u_{s,t}(X_t))\eqsp .
    \end{align*}
After replacing $u$ by $u^\theta$ in the previous equality, we obtain the ESD target \eqref{eq:general_esd_loss} from the right term.

\noindent\textbf{Shortcut setting.} By definition of the Flow Map given in \eqref{eq:flowmap_def}, we have for any time $r\in (s,t)$
    \begin{align*}
        X_{s,t}(X_t) = X_{s,r}(X_{r,t}(X_t)) \eqsp .
    \end{align*}
When substituting the parametrization \eqref{eq:param_euler} in the previous equality, we obtain the \textit{Shortcut characterization}
    \begin{align*}
        &X_t - (t-s) u_{s,t}(X_t) = \{X_t-(t-r)u_{r,t}(X_t)\}-(r-s)u_{s,r}(X_{r,t}(X_t))\\
        \iff & u_{s,t}(X_t) = \frac{t-r}{t-s}u_{r,t}(X_t) + \frac{r-s}{t-s}u_{s,r}(X_{r,t}(X_t)) \eqsp .
    \end{align*}
After replacing $u$ by $u^\theta$ in the previous equality, we obtain the SSD target \eqref{eq:general_ssd_loss} from the right term.
\end{proof}

\paragraph{Simplifications with the standard interpolant.} Now assume that we have $\mathrm{I}_t= (1-t) X_0 + t X_1$. To obtain the objective functions presented in \Cref{subsec:flowmap}, we proceed to the following simplifications: (a) taking inspiration from the conditional trick in FM, we replace each occurrence of the \textit{marginal} velocity field ($u^\theta_{s,s}$ in LSD loss, $u^\theta_{t,t}$ in ESD loss) by the \textit{conditional} velocity field, exactly defined as $v_{\text{FM}}^{\text{target}}=X_1-X_0$, which has the computational advantage to be constant over the whole interpolant, (b) following \cite{frans2025one}, we deterministically set $q(r|s,t)=\delta_{(s+t)/2}(r)$ in SSD formulation.

\subsection{Generalization of CFG to the Flow Map setting} \label{subsec:flowmap_general_cfg}

To incorporate CFG into Flow Map models, we suggest to replace the target velocity field $v_t$ by its CFG counterpart \eqref{eq:cfg_standard}, where the class-unconditional velocity $v_t(\cdot|\emptyset)$ is legitimately substituted by $u_{t,t}(\cdot|\emptyset)$. This naturally leads to the CFG-based loss for the FM setting detailed in \eqref{eq:cfg_loss}, which is unchanged in the general SI setting (recalling that we now have $v_{\text{FM}}^{\text{target}}=\dot{\mathrm{I}}_t$). With this paradigm, the SD targets defined in \Cref{subsec:flowmap_general} naturally turn into the following CFG-based targets
\begin{enumerate}
    \item CFG-based \textit{Lagrangian} SD (cfg-LSD) loss
    \begin{align*}
        & u_{\text{cfg-LSD}}^{\text{target}} =  v^{\text{target}}_{\text{cfg-FM}}  + (t-s) \partial_s u^\theta_{s,t}(\mathrm{I}_t|\mathbf{c}) \eqsp, \\
        &\quad v^{\text{target}}_{\text{cfg-FM}} = w\, u^\theta_{s,s}\left(\mathrm{I}_t -(t-s)u^\theta_{s,t}(\mathrm{I}_t|\mathbf{c})\mid\mathbf{c}\right) + (1-w) u^\theta_{s,s}\left(\mathrm{I}_t -(t-s)u^\theta_{s,t}(\mathrm{I}_t|\emptyset )\mid\emptyset\right) \eqsp,
    \end{align*}
    \item CFG-based \textit{Eulerian} SD (cfg-ESD) loss
    \begin{align*}
        &u_{\text{cfg-ESD}}^{\text{target}} = v^{\text{target}}_{\text{cfg-FM}}-(t-s)\left(\nabla u^\theta_{s,t}(\mathrm{I}_t|\mathbf{c}) v^{\text{target}}_{\text{cfg-FM}} + \partial_t u^\theta_{s,t}(\mathrm{I}_t|\mathbf{c})\right) \eqsp ,\\
        &\quad v^{\text{target}}_{\text{cfg-FM}} = w\, u^\theta_{t,t}(\mathrm{I}_t|\mathbf{c}) + (1-w) u^\theta_{t,t}(\mathrm{I}_t|\emptyset) \eqsp ,
    \end{align*}
    \item CFG-based \textit{Shortcut} SD (cfg-SSD) loss
    \begin{align*}
        u_{\text{cfg-SSD}}^{\text{target}}= \frac{t-r}{t-s}u^\theta_{r,t}(\mathrm{I}_t|\mathbf{c}) + \frac{r-s}{t-s}u^\theta_{s,r}\left(\mathrm{I}_t -(t-r) u^{\theta}_{r,t}(\mathrm{I}_t|\mathbf{c}) \mid \mathbf{c}\right)\eqsp , \eqsp r\sim q(r|t,s) \eqsp.
    \end{align*}
\end{enumerate}

\paragraph{Simplifications with the standard interpolant.} Now assume that we have $\mathrm{I}_t= (1-t) X_0 + t X_1$. To obtain the CFG-based SD objectives presented in \Cref{subsection:cfg_flowmap}, we apply the same simplifications as in the unconditional setting: in particular, we replace each occurrence of the \emph{class-conditional marginal} velocity field in LSD and ESD settings by its conditional counterpart $v_{\text{FM}}^{\text{target}}=X_1-X_0$, thus avoiding extra model evaluations, and set $q(r|s,t)=\delta_{(s+t)/2}(r)$ in SSD setting.

\section{Implementation details} \label{sec:implem_details}

\subsection{General settings}

\paragraph{Data.} We assemble a collection of publicly available, free-to-use images to construct the LR–HR training dataset $\mathcal{D}$. To generate synthetic LR images from HR ground truth, we adapt the widely used degradation pipeline of \cite{wang2021realesrgan}, applying the following operations in sequence:
\begin{enumerate}
\item \emph{Moderate blurring}: applied with $80\%$ probability;
\item \emph{Downscaling} (core operation): applied systematically using a random scale $s_{\text{down}} \sim \mathrm{U}(0.1, 1)$ and a randomly chosen interpolation mode (nearest, area, bilinear, or bicubic);
\item \emph{Addition of low-level noise}: Gaussian noise with $50\%$ probability, otherwise Poisson noise;
\item \emph{JPEG compression}: applied systematically with the default compression level from \cite{wang2021realesrgan}.
\end{enumerate}
To match the spatial resolution of the HR image and define a valid interpolant for the Flow Map formulation, the degraded output is resized back using a random interpolation mode, followed by additional blurring and pixel-value clamping. Both HR and LR images are then rescaled to the range $[-1, 1]$. Since we operate in latent space to leverage large pre-trained diffusion models, both HR and synthetically generated LR images are embedded using the pre-trained variational encoder from \citep{podell2024sdxl}.

To support training across multiple resolutions, we adopt a bucketing strategy \citep{podell2024sdxl} that accommodates varying aspect ratios and image sizes. The specific configuration depends on the scalability of each Flow Map variant:
\begin{itemize}[wide, labelindent=0pt]
\item \textit{SSD setting}: pixel budgets of $[512^2, 1024^2]$ sampled with probabilities $[0.2, 0.8]$ and batch sizes $[4, 2]$, respectively, to fit within a single H100 GPU; aspect ratios are sampled in $[1/4, 4]$.
\item \textit{LSD and ESD settings}: pixel budgets of $[256^2, 512^2]$ with the same sampling probabilities and batch sizes as in the SDD setting. Higher resolutions could not be accommodated due to memory constraints during training, which could not be alleviated by multi-GPU sharding owing to software limitations imposed by the use of the $\operatorname{jvp}$ package. Aspect ratios are again sampled in $[1/4, 4]$.
\end{itemize}

\paragraph{Architectures.} In the FlowMapSR framework described in \Cref{sec:method}, we employ three neural network components:
\begin{enumerate}[wide, labelindent=0pt, label=(\Alph*)]
\vspace{-0.2cm}
\item \emph{Flow Map model} $u^\theta_{s,t}$.
This component is implemented as a UNet with an architecture inspired by the SDXL text-to-image diffusion model \citep{podell2024sdxl}. Conditioning is handled via projection-based embeddings. Specifically, to evaluate the model at a timestep pair $(s, t)$, we apply positional encodings to each input, using 256 channels per timestep by default. In addition, conditioning on positive or negative text prompts is achieved through a frozen CLIP text encoder, producing embeddings of dimension 1280. Overall, the resulting model contains approximately $2.5$B parameters.
\item \emph{Dynamic weighting function} $\lambda^\psi_{s,t}$.  
This component is implemented as a lightweight neural network. It consists of positional embeddings of $s$ and $t$ (256 channels each), whose outputs are concatenated into a 512-dimensional vector and passed through a two-layer MLP ($512 \to 1$ with SiLU activation, followed by $1 \to 1$). The final layer is initialized to zero using Kaiming uniform initialization, ensuring that the initial weighting is constant and equal to 1 across all timestep pairs. As a result, this network is extremely compact, with fewer than $1$k parameters.

\item \emph{Discriminator} $\mathbf{D}^\phi$.  
It follows a patch-based architecture in the style of Pix2Pix \citep{isola2017image}. It comprises four Conv2d layers with 64, 128, 256, and 512 channels, respectively. Group normalization is applied after each layer except the first, followed by a SiLU activation. A final Conv2d layer with a single output channel produces a patch-wise output of spatial size $(w/8 - 2, h/8 - 2)$ for an input of size $(w, h)$. The discriminator contains approximately $2.8$M parameters.
\end{enumerate}

\subsection{Training procedure}

\paragraph{Initialization of unconditional FlowMapSR.}
For unconditional models, we initialize the network using the Latent Bridge Matching (LBM) image deblurring checkpoint from \cite{chadebec2025lbm}, which provides a strong initialization for image restoration. Although the LBM interpolant is defined as a Brownian bridge—i.e., a \textit{stochastic} variant of the standard FM interpolant with added Gaussian noise scaled by $\sigma_t=\sigma\sqrt{t(1-t)}$—this initialization remains appropriate in our setting, since the noise level is negligible ($\sigma=5\times10^{-3}$). As the LBM model is unconditional, we load only the shared parameters, leaving class-embedding weights randomly initialized.
\vspace{-0.2cm}
\paragraph{Initialization of CFG-based FlowMapSR.}
CFG-based models are initialized from their corresponding unconditional FlowMapSR counterparts, previously trained without guidance. As before, only shared parameters are loaded, while class-embedding weights are initialized randomly.
\vspace{-0.2cm}
\paragraph{Time-dependent regularization with perceptual loss.}
We set the time-dependent perceptual weight as $\lambda^{\text{LPIPS}}_s = 5\times\exp(-4s)$, which we found to balance effective pixel-level regularization with minimal bias toward the FM-SD objective. To reduce memory usage, we follow the random cropping strategy of \cite{chadebec2025lbm}, computing the $\operatorname{LPIPS}$ loss on randomly sampled patches from $X_0$ and $\hat{X}_0$. In all experiments, the maximum patch size is fixed to $512^2$.
\vspace{-0.2cm}
\paragraph{Discrete timestep sampling.}
Let $\{t_k\}_{k=0}^K$ denote a discretization of $[0,1]$ into $K$ uniform sub-intervals. We set $d_{\max}=7$, corresponding to a finest discretization of $2^7+1=129$ timesteps. The joint sampling distribution $\bar{q}(s,t)$ is defined as follows, depending on the loss type:
\begin{itemize}[wide, labelindent=0pt]
\vspace{-0.2cm}
\item \textit{FM loss.} With $75\%$ probability, we set $d=d_{\max}$, sample $k$ uniformly from $\{0,\ldots,K_d\}$, and take $s=t=t_k$.
\item \textit{SD loss (LSD/ESD).} With $25\%$ probability, we sample $d$ uniformly from $\{0,\ldots,d_{\max}\}$, then sample $k$ uniformly from $\{1,\ldots,K_d\}$ and $k'$ uniformly from $\{0,\ldots,k\}$, setting $(s,t)=(t_{k'},t_k)$.
\item \textit{SD loss (SSD).} With $25\%$ probability, we sample $d$ uniformly from $\{0,\ldots,d_{\max}-1\}$, sample $k$ uniformly from $\{1,\ldots,K_d\}$, and set $(s,t)=(t_{k-1},t_k)$. While more restrictive than the LSD/ESD setting, this choice simplifies the definition of the midpoint $r=(s+t)/2$ and matches the protocol of \cite{frans2025one}.
\end{itemize}
\vspace{-0.2cm}
\paragraph{Positive–negative prompting guidance \& LoRA-based adversarial fine-tuning.}
Following standard CFG practice, we sample the positive prompt $\mathbf{c}^{\text{pos}}$ with $90\%$ probability and the negative prompt $\mathbf{c}^{\text{neg}}$ with $10\%$ probability. We set $\lambda^{\text{adv}}=0.1$ and fix the LoRA rank to 64 across all experiments.
\vspace{-0.2cm}
\paragraph{Optimization details.}
Training instances of FlowMapSR begins with a pure FM warm-start phase of $5$k iterations, followed by $5$k iterations of unconditional self-distillation and an additional $3$k iterations with CFG. Adversarial fine-tuning then proceeds for $2.5$k to $4$k iterations, depending on RPGAN convergence. The effective batch size is fixed to 256 using gradient accumulation. We use AdamW with default hyperparameters, setting the learning rate to $4\times10^{-5}$ for the Flow Map model and $10^{-4}$ for the discriminator. Since the initial LBM checkpoint is already well aligned with our framework, we do not employ EMA, which we found to degrade performance.

\subsection{Unsuccessful enhancement strategies}

In preliminary experiments, we explored several alternative strategies to further enhance SR performance. However, none of the approaches described below led to measurable improvements over our final configuration.
\begin{enumerate}[wide, labelindent=0pt, label=(\alph*)]
\vspace{-0.2cm}
\item \emph{Conditioning on the upscaling factor $s_{\text{up}}$.} We investigated explicitly conditioning the Flow Map model on the target upscaling factor using positional encodings with appropriate scaling, but it did not yield any noticeable performance gains.
\vspace{-0.1cm}
\item \emph{Fourier coefficient matching loss.} Inspired by \citep{wu2025onestepdiffusionbasedrealworldimage}, we evaluated the use of Fourier coefficient matching as an additional pixel-level regularization term, using the same coefficient settings as for the LPIPS loss. In our experiments, this approach did not improve output sharpness or visual quality.
\vspace{-0.1cm}
\item \emph{Pixel-space RPGAN fine-tuning.} To exhaustively explore adversarial training options, we also experimented with aligning pixel-space distributions after CFG training. To mitigate memory constraints, we adopted the same random cropping strategy as for LPIPS regularization, evaluating the discriminator on randomly sampled patches from $\hat{X}^{\text{adv}}_0$ and $X_0$, with a maximum patch size of $512^2$. This approach consistently produced inferior results compared to our default latent-space adversarial fine-tuning. Representative qualitative examples are shown in \Cref{fig:all_gan}.
\end{enumerate}
\vspace{-0.4cm}
\section{Additional results} \label{sec:additional_results}
\vspace{-0.2cm}

\paragraph{Additional comparison with competing methods.}

\begin{figure}[h!]
    \centering
    \includegraphics[width=\linewidth]{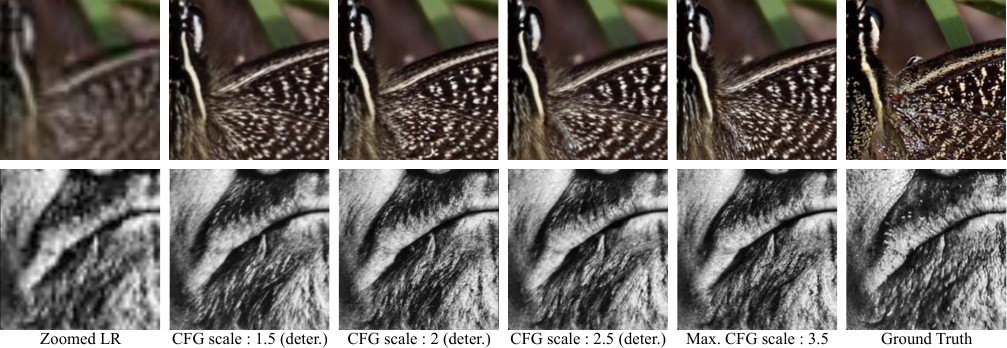}
    \vspace{-0.7cm}
    \caption{\textbf{Qualitative comparison of CFG-enhanced Shortcut variants of FlowMapSR for $\times 4$ upscaling (deterministic vs stochastic).}
We compare several deterministic choices of the guidance scale $w$ with the default stochastic CFG configuration ($w_{\max}=3.5$). Adversarial fine-tuning is not applied in these comparisons. The LR images are taken from the Div2K-Val dataset.}
\vspace{-2cm}
    \label{fig:cfg_deter}
\end{figure}

In this section, we complement the main results of \Cref{sec:expes}. To demonstrate the consistency of FlowMapSR across upscaling factors, we report quantitative results on DIV2K-Val for $\times2$ SR in \Cref{table:upscaling_2}, with qualitative examples in \Cref{fig:x2_div2k}. These results align with those observed for $\times4$ and $\times8$ upscaling in \Cref{sec:expes}. We further present qualitative comparisons for $\times4$ upscaling on DIV2K-Val (\Cref{fig:x4_div2k}) and RealSR (\Cref{fig:x4_realsr}), as well as for $\times8$ upscaling on DIV2K-Val (\Cref{fig:x8_div2k}) and RealSR (\Cref{fig:x8_realsr}). Real-world LR inputs from the RealSet65 dataset are also included in \Cref{fig:real_world_app} for both $\times4$ and $\times8$ SR. Across all settings, FlowMapSR consistently recovers fine-grained structures aligned with HR references, produces lifelike textures, particularly for human and animal details (e.g., rows 5–8 in \Cref{fig:x4_div2k}, rows 3–4 in \Cref{fig:real_world_app}, and rows 2, 3, and 7 in \Cref{fig:x8_div2k}), and avoids visually disruptive artifacts. Moreover, FlowMapSR more faithfully preserves realistic depth of field, where competing methods often fail (e.g., row 7 in \Cref{fig:x4_div2k}, rows 2 and 7 in \Cref{fig:x4_realsr}, and row 8 in \Cref{fig:x8_div2k}).
\vspace{-0.2cm}
\paragraph{Ablation studies.}
We further provide complementary qualitative analyses to illustrate the effect of key design choices in FlowMapSR. In \Cref{fig:cfg_deter}, we compare \emph{deterministic} CFG training with the \emph{stochastic} CFG strategy proposed in this work, observing reduced sharpness and increased artifacts under deterministic guidance. In \Cref{fig:app_lsd_esd}, we report additional examples for the \emph{Lagrangian} and \emph{Eulerian} FlowMapSR variants, with and without CFG: the \emph{Lagrangian} formulation deteriorates rapidly as CFG strength increases, while the \emph{Eulerian} variant remains more stable but struggles to produce visually compelling results. In \Cref{fig:all_gan}, we analyze the impact of adversarial fine-tuning in the \emph{Shortcut} formulation, comparing latent-space and pixel-space RPGAN approaches under both unconditional and CFG-based training. Latent-space adversarial fine-tuning consistently yields superior visual quality, and its benefits are significantly amplified when combined with CFG. These observations motivate our final design choice, where adversarial fine-tuning is applied jointly with CFG rather than in isolation. Finally, we provide additional qualitative examples for the ablation studies on the number of inference steps (\Cref{fig:app_num_steps}) and the LoRA scale (\Cref{fig:app_lora_scale}), which further confirm the trends discussed in \Cref{sec:expes}.

\newpage
\begin{figure}[h!]
\vspace{-1cm}
    \centering
    \includegraphics[width=\linewidth]{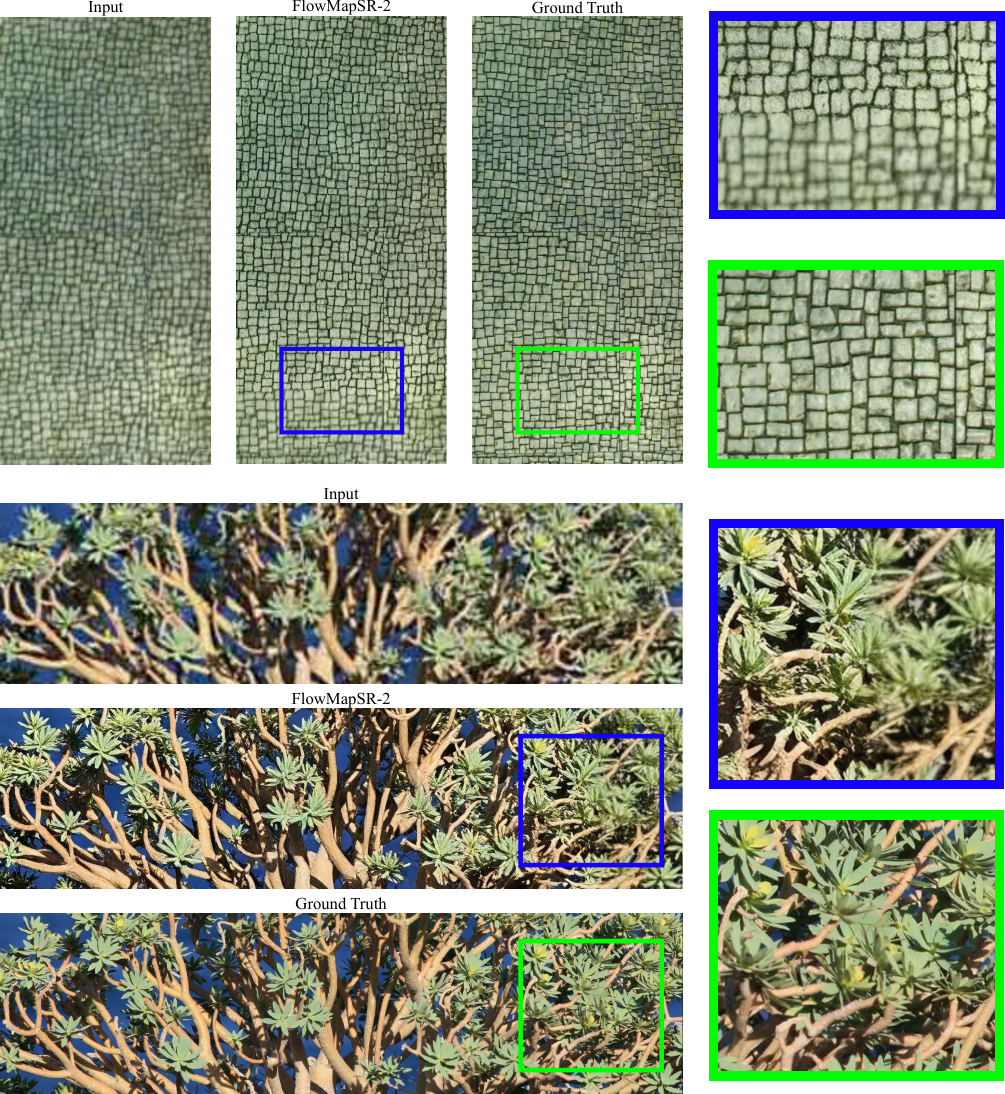}
    \vspace{-0.5cm}
    \caption{\textbf{Examples of tiling failure in FlowMapSR for $\times 4$ upscaling.}
The LR inputs are drawn from the DIV2K-Val dataset. For HR images, we display the same zoomed region (\tcmb{blue} for FlowMapSR, \tcmv{green} for ground truth).}
    \label{fig:app_tiling}
\end{figure}

\paragraph{Visual limitation of FlowMapSR.} We observe in some cases that the Gaussian tiling strategy used in FlowMapSR to generate HR images with resolution exceeding $1024 \times 1024$ introduces visible artefacts, manifested as blurry rectangular regions near image boundaries, see \Cref{fig:app_tiling}. We conjecture that this limitation may contribute to the inferior performance observed on non-reference evaluation metrics in \Cref{table:upscaling_4_8}, as such artefacts degrade image sharpness and perceptual realism. Addressing and improving the tiling procedure is left for future work.

\newpage

\begin{table}[h!]
\vspace{-1cm}
\caption{ \textbf{Quantitative comparison between FlowMapSR and competing SR methods ($\times 2$ upscaling).} The best and second best results are highlighted in \textbf{bold} and \underline{underlined}, respectively.}
\vspace{-0.3cm}
\label{table:upscaling_2}
\begin{center}
\resizebox{\textwidth}{!}{%
\begin{tabular}{r|l|ccccc|cccc}
\toprule
\multirow{2}{*}{Dataset ($s_{\text{up}}$)} & \multirow{2}{*}{Method} & \multicolumn{5}{c|}{\underline{Reference metrics}} & \multicolumn{4}{c}{\underline{Non-reference metrics}}\\[0.4em]
 & & PSNR$\uparrow$ & SSIM$\uparrow$ & LPIPS$\downarrow$ & DISTS$\downarrow$ & FID$\downarrow$ & NIQE$\downarrow$ & MANIQA$\uparrow$ & MUSIQ$\uparrow$ & CLIPIQA$\uparrow$\\
\midrule
\multirow{4}{*}{DIV2K-Val ($\times 2$)} & Real-ESRGAN & \bf 25.16 & \bf 0.7829 & \underline{0.2123} & 0.1050 & 14.75 & 3.855 & \underline{0.4122} & \underline{0.6450} & \underline{0.5515} \\
 & OSEDiff-1 & 22.72& 0.6931 & 0.2363 & 0.1063 & 17.78 & \underline{3.386} & \bf 0.4533 & \bf 0.6894 & \bf 0.6183 \\
  & S3Diff-1 & 22.39 & 0.6748 & 0.2316 & \underline{0.1034} & \underline{10.97} & 3.715 & 0.3700 &  0.6175 & 0.5045\\
  & \cellred{FlowMapSR-2} & \cellred{\underline{23.82}} & \cellred{\underline{0.6989}} & \cellred{\bf 0.1999} & \cellred{\bf 0.0770} & \cellred{\bf 7.16} & \cellred{\bf 2.904} & \cellred{0.3892} & \cellred{0.6390} & \cellred{0.5356}\\
\bottomrule
\end{tabular}
}
\end{center}
\end{table}

\begin{figure}[h!]
    \centering
    \includegraphics[width=\linewidth]{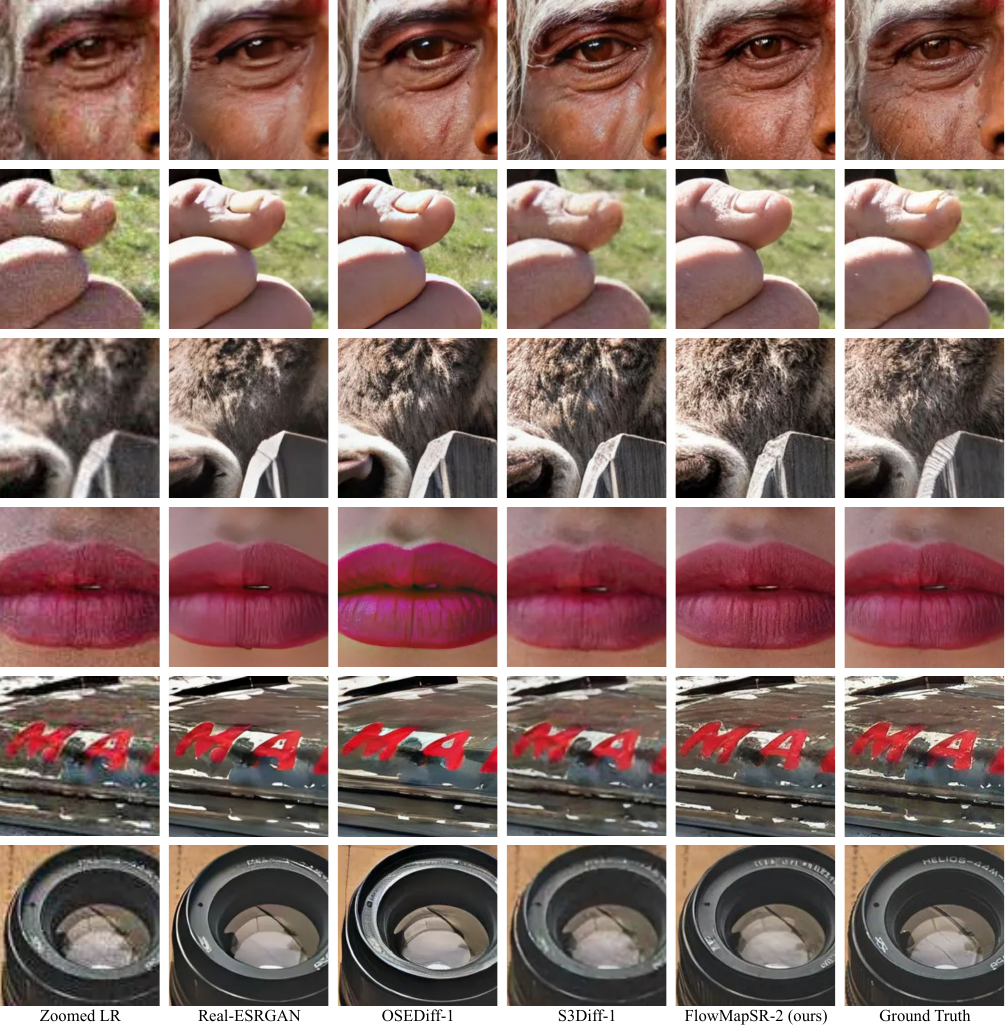}
    \vspace{-0.7cm}
    \caption{\textbf{Qualitative comparison between FlowMapSR and competing SR methods for \underline{$\times 2$ upscaling} on DIV2K-Val.} Visually, FlowMapSR more faithfully recovers fine details, textures and depth of field while maintaining high overall image quality with no noticeable artifacts.}
    \vspace{-2cm}
    \label{fig:x2_div2k}
\end{figure}

\newpage

\begin{figure}
\vspace{-1.4cm}
    \centering
    \includegraphics[width=\linewidth]{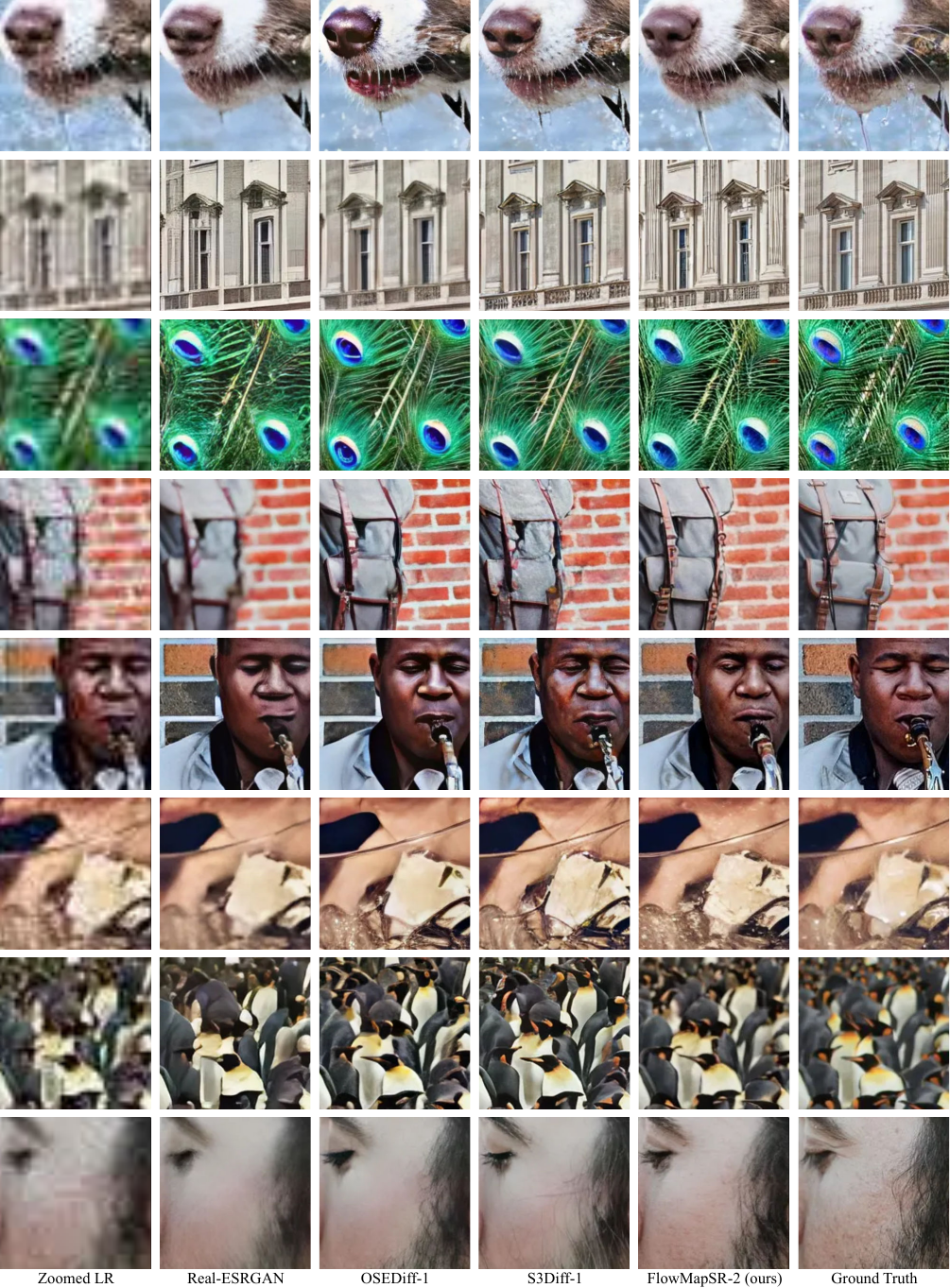}
    \vspace{-0.5cm}
    \caption{\textbf{Qualitative comparison between FlowMapSR and competing SR methods for \underline{$\times 4$ upscaling} on DIV2K-Val.} This is complementary to \Cref{fig:x4_x8_upscaling}.}
    \label{fig:x4_div2k}
\end{figure}

\newpage

\begin{figure}
    \centering
    \includegraphics[width=\linewidth]{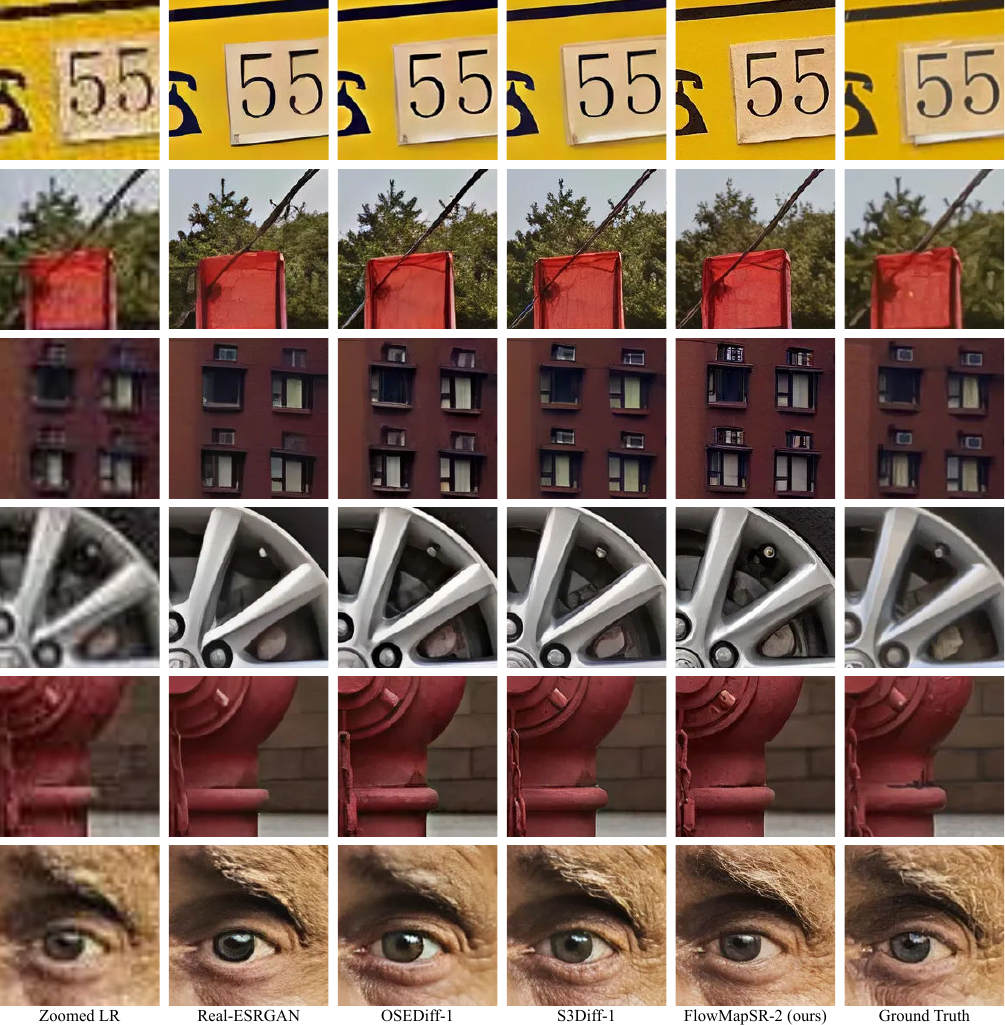}
    \caption{\textbf{Qualitative comparison between FlowMapSR and competing SR methods for \underline{$\times 4$ upscaling} on RealSR.} This is complementary to \Cref{fig:x4_x8_upscaling}.}
    \label{fig:x4_realsr}
\end{figure}

\newpage

\begin{figure}
\vspace{-1.4cm}
    \centering
    \includegraphics[width=\linewidth]{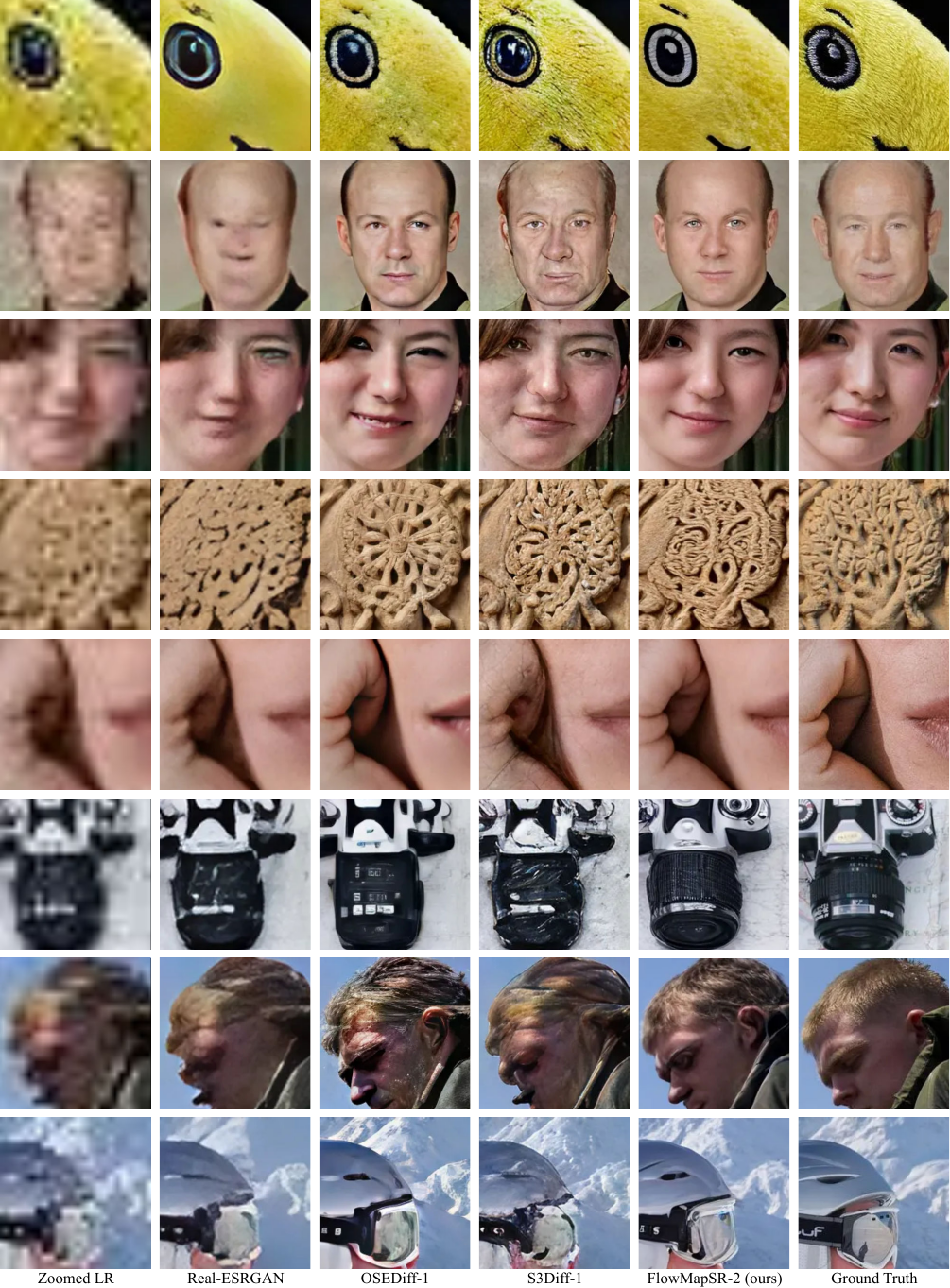}
    \vspace{-0.5cm}
    \caption{\textbf{Qualitative comparison between FlowMapSR and competing SR methods for \underline{$\times 8$ upscaling} on DIV2K-Val.} This is complementary to \Cref{fig:x4_x8_upscaling}.}
    \label{fig:x8_div2k}
\end{figure}

\newpage

\begin{figure}
    \centering
    \includegraphics[width=\linewidth]{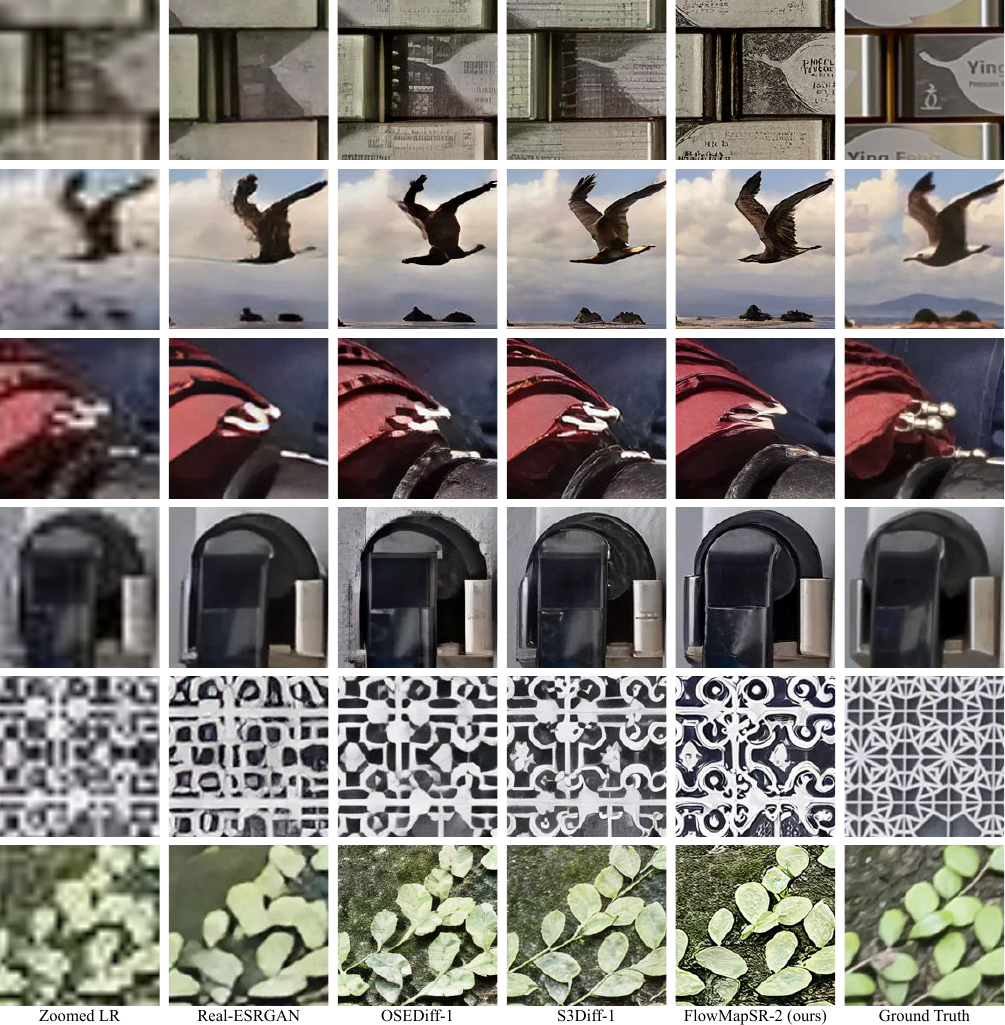}
    \caption{\textbf{Qualitative comparison between FlowMapSR and competing SR methods for \underline{$\times 8$ upscaling} on RealSR.} This is complementary to \Cref{fig:x4_x8_upscaling}.}
    \label{fig:x8_realsr}
\end{figure}

\newpage

\begin{figure}
\vspace{-1.4cm}
    \centering
    \includegraphics[width=\linewidth]{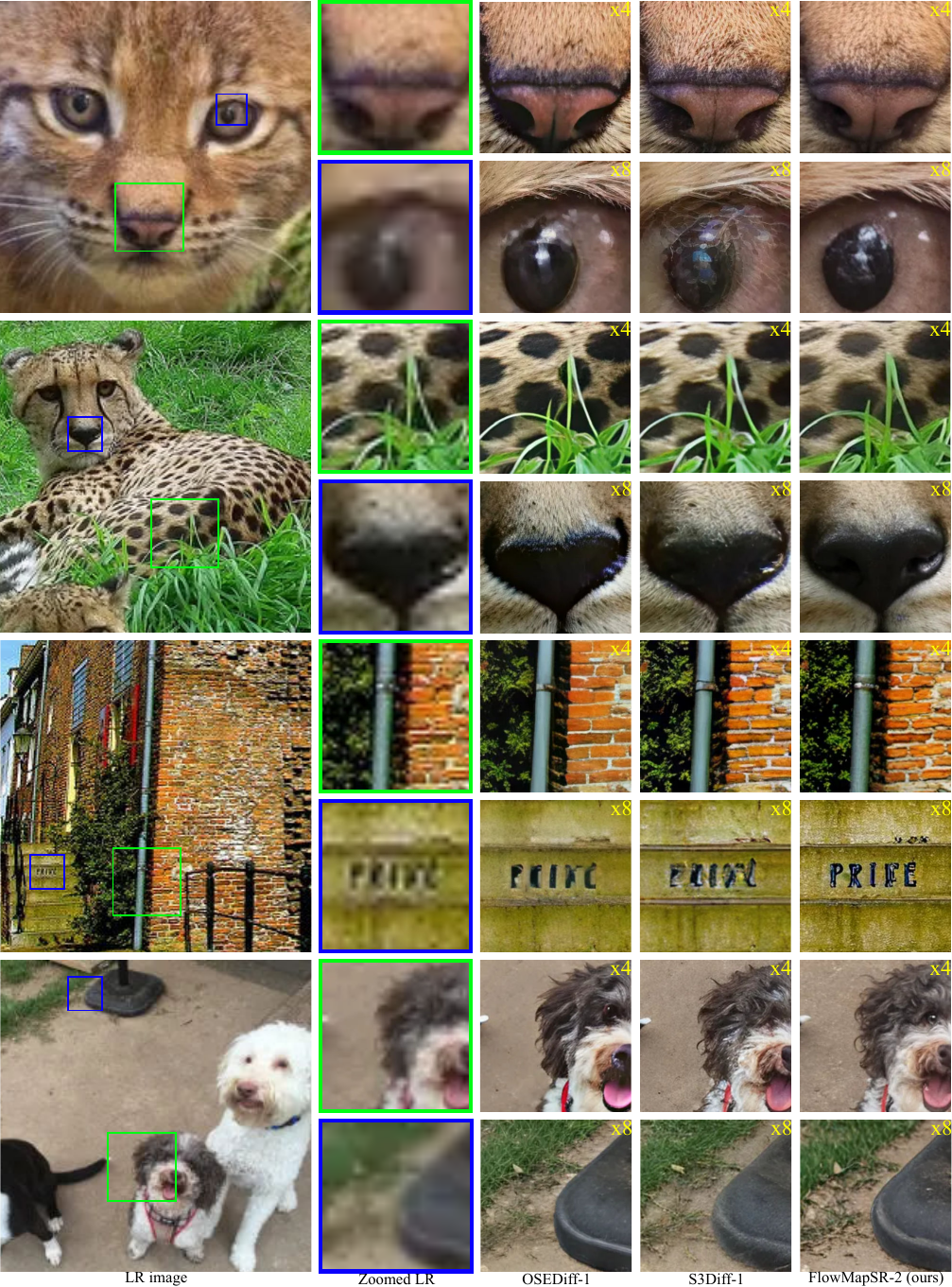}
    \vspace{-0.6cm}
    \caption{\textbf{Qualitative comparison between FlowMapSR and competing diffusion-based SR methods on real-world LR inputs for $\times 4$ and $\times 8$ upscaling.} The LR images are taken from the RealSet65 dataset. This is complementary to \Cref{fig:real_main}.}
    \label{fig:real_world_app}
\end{figure}

\newpage

\begin{figure}
\vspace{-1.4cm}
    \centering
    \includegraphics[width=\linewidth]{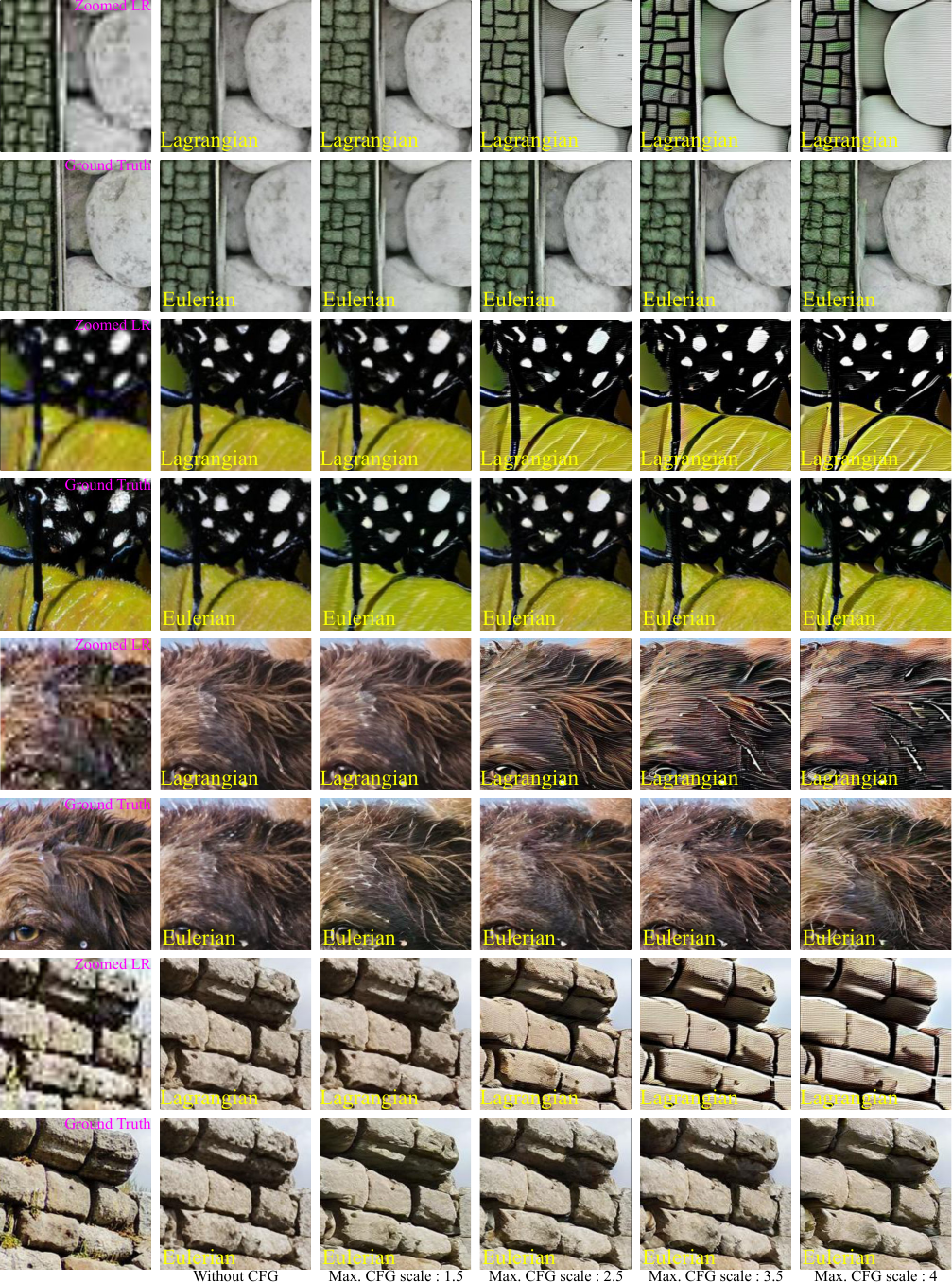}
    \vspace{-0.6cm}
    \caption{\textbf{Qualitative comparison of positive-negative CFG enhancement applied to LSD- and ESD-based FlowMapSR for $\times 4$ and $\times 8$ upscaling.} The LR images are taken from the Div2K-Val dataset. This is complementary to \Cref{fig:cfg_impact_all_sd}.}
    \label{fig:app_lsd_esd}
\end{figure}

\newpage

\begin{figure}
    \centering
    \includegraphics[width=\linewidth]{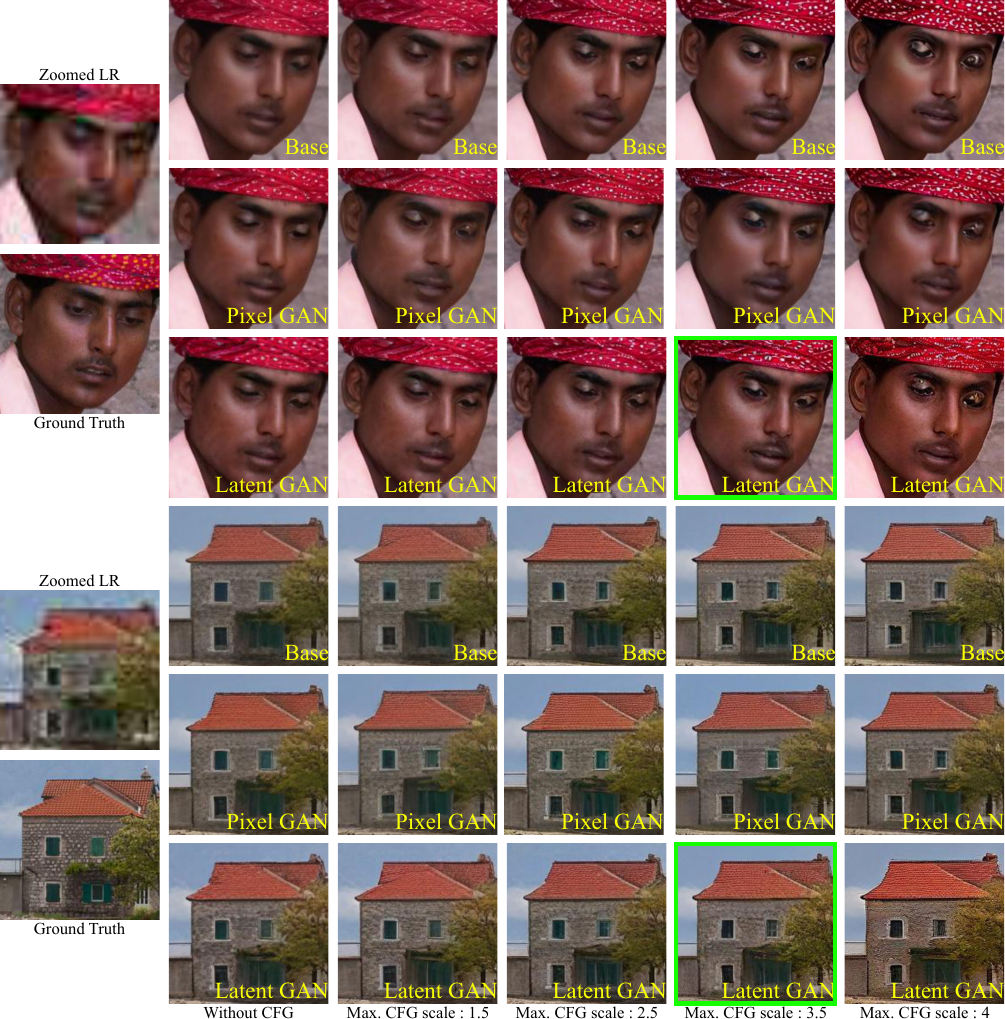}
    \vspace{-0.5cm}
    \caption{\textbf{Qualitative comparison of adversarially finetuned Shortcut variants of FlowMapSR for $\times 4$ upscaling.} The \tcmv{green} setting ($w_{\max}=3.5$ \& latent RPGAN) is the default configuration used in our main experiments. The LR images are taken from the Div2K-Val dataset. For each LR–HR pair, increasing the CFG scale alone (first row) already leads to noticeably sharper results, reflecting improved perceptual quality. When adversarial fine-tuning is additionally applied, texture fidelity and sharpness further improve. This effect is especially pronounced when fine-tuning is performed in latent space (third row), whereas pixel-space fine-tuning yields only marginal gains (second row). Overall, combining CFG with latent adversarial fine-tuning produces superior visual results compared to using solely CFG or adversarial fine-tuning on the standard FlowMapSR model.}
    \label{fig:all_gan}
\end{figure}

\newpage

\begin{figure}[h!]
    \centering
    \includegraphics[width=\linewidth]{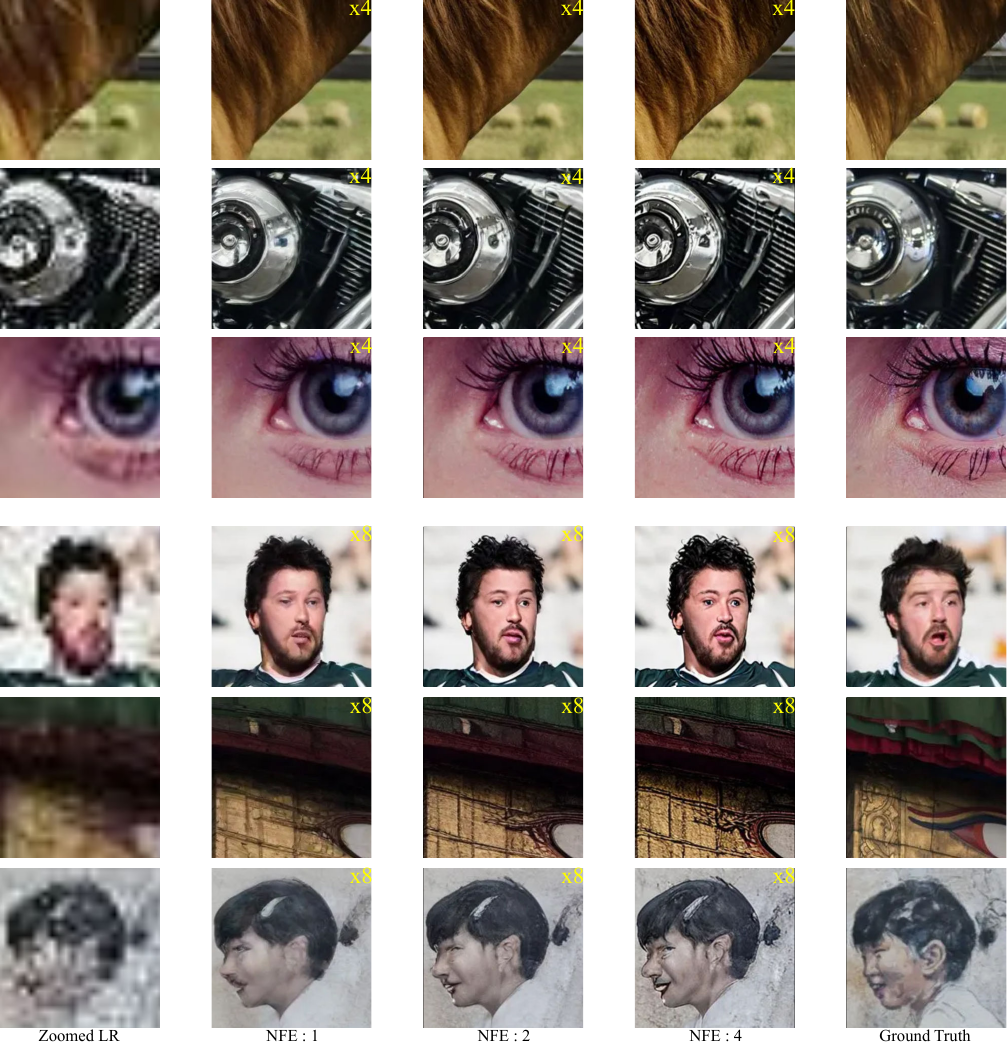}
    \vspace{-0.5cm}
    \caption{\textbf{Qualitative comparison of FlowMapSR with varying numbers of inference steps (NFE) for $\times 4$ (rows 1–3) and $\times 8$ (rows 4–6) upscaling.}
The LR inputs are drawn from the DIV2K-Val dataset. This is complementary to \Cref{fig:ablation_num_steps}.}
    \label{fig:app_num_steps}
\end{figure}

\begin{figure}[h!]
    \centering
    \includegraphics[width=\linewidth]{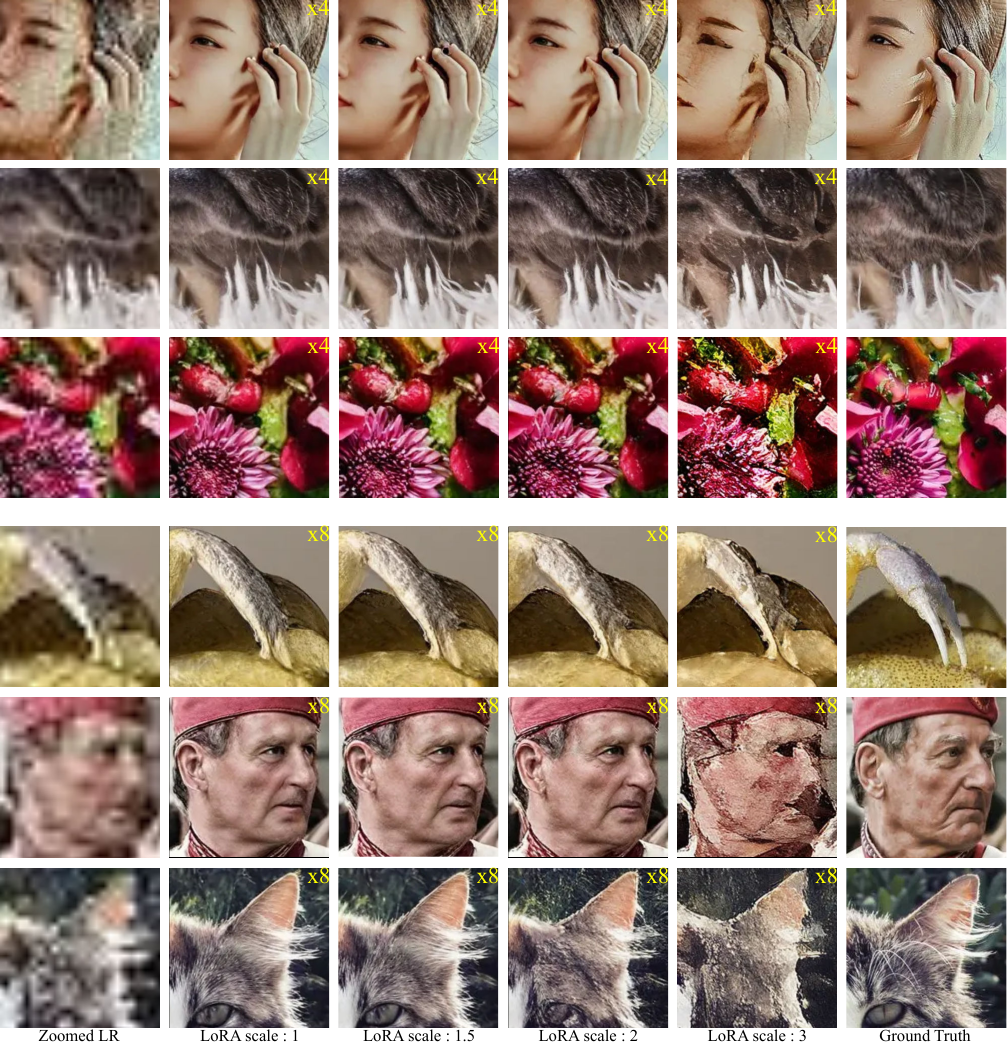}
    \vspace{-0.5cm}
    \caption{\textbf{Qualitative comparison of FlowMapSR with varying LoRA scale for $\times 4$ (rows 1–3) and $\times 8$ (rows 4–6) upscaling.}
The LR inputs are drawn from the DIV2K-Val dataset. This is complementary to \Cref{fig:ablation_lora_scale}.}
    \label{fig:app_lora_scale}
\end{figure}

\end{document}